\documentclass[12pt]{article}

\pdfoutput=1
\usepackage{amssymb,textcomp,amsmath,graphicx}
\DeclareMathAlphabet{\mathcal}{OMS}{cmsy}{m}{n}
\long\def\symbolfootnote[#1]#2{\begingroup
\def\thefootnote{\fnsymbol{footnote}}
\footnote[#1]{#2}\endgroup}
\def\lsim{\:\raisebox{-0.75ex}{$\stackrel{\textstyle<}{\sim}$}\:}
\def\gsim{\:\raisebox{-0.75ex}{$\stackrel{\textstyle>}{\sim}$}\:}

\newcommand{\be}{\begin{equation}} \newcommand{\ee}{\end{equation}}
\newcommand{\bea}{\begin{eqnarray}} \newcommand{\eea}{\end{eqnarray}}

\newcommand{\nn}{\nonumber}
\newcommand{\mpl}{M_{\rm P}}
\newcommand{\lp}{\left(}
\newcommand{\rp}{\right)}

\newcommand{\ie}{{\em i.e.}}
\newcommand{\eg}{{\em e.g.}}

\newcommand{\etal}{\emph{et\,al.~}}
\newcommand{\p}{{\it Planck}}

\numberwithin{equation}{section}

\begin{document}
\begin{center}
\Large{\bf Modulated Inflation Models and Primordial Black Holes}
\end{center}
\begin{center}
\large{Encieh Erfani$^{*}$}
\end{center}
\begin{center}
\textit{School of Physics, Institute for Research in Fundamental Sciences (IPM),\\ P. Code. 19538-33511, Tehran, Iran}
\end{center}

\date{}

\symbolfootnote[0]{$^{*}$eerfani@ipm.ir}

\begin{abstract}
For the first time, the running of the running of the spectral index $\beta_s$ has been measured by {\it Planck}. This parameter is crucial for dark matter Primordial Black Holes (PBHs) formation. We study the PBHs formation in inflation potentials with superimposed periodic oscillations. We show that the value of $\beta_s$ suggested by the recent cosmic microwave background data is easily explained in these models. As a by-product, we also show that the nonproduction of (long--lived) PBHs puts a stronger upper bound on $\beta_s$.
\end{abstract}

\newpage

\section{Introduction}
Recently, the inflationary paradigm has become an essential part of our description of the Universe which solves cosmological puzzles of the hot big bang model such as horizon and flatness of the Universe \cite{Guth}. Moreover, it provides a natural mechanism for the generation of the spectrum of primordial fluctuations in the early universe which are in good agreement with observations \cite{Planck1, Planck2}. These fluctuations can also explain the generation of inhomogeneities that can be seen in our Universe, from the cosmic microwave background (CMB) anisotropies to the large scale structure (LSS) in the form of galaxies and clusters \cite{Lyth_book}.

Albeit these successes, the specific details of the inflation scenario are still unknown, since many different mechanisms and fields may generate a phase of accelerated cosmic expansion. For simplicity, people consider a scalar field (called inflaton) with a simple form for its potential which generates inflation. In addition, many models of inflation \cite{review} make accurate predictions for observations to discriminate between various models. One such prediction concerns the possibility of primordial black holes (PBHs) formation.

It was shown that the spectrum of primordial fluctuations could lead to the production of PBHs \cite{Hawking, Carr}, and for this mechanism, one typically needs a ``blue" spectrum; \ie, the value of the spectral index in the length scale of PBHs formation, which is much larger than the scales probed by the CMB and the LSS data, should be larger than $1$ \cite{{PBH_formation, Encieh1, Encieh2}}. Because of this discrepancy in scales, PBH formation becomes a unique probe of cosmological inflation; in particular, the constraint that PBHs do not overclose our Universe has been used to limit the power spectrum at small length scales, which, in turn allows
to constrain inflationary models \cite{PBH_constraints}.

For comparing inflation models by observation, the relevant parameters are the scalar spectral index $n_s$, the ratio of tensor to scalar fluctuations $r$, the running of the scalar spectral index $\alpha_s$, and the ``running of running of the spectral index" $\beta_s$ which has been measured for the first time by {\it Planck} \cite{Planck2}. In Ref.~\cite{Encieh1} the importance of this parameter for the production of (long-lived) PBHs as a candidate for dark matter (DM) has been studied.

The CMB observations confirm that the spectral index $n_s$ has red tilt; hence, for the PBHs formation, one needs a positive value for the running of the spectral index $\alpha_s$. However, the observational data indicates a large negative running of the spectral index of order $10^{-2}$ \cite{Planck1, Planck2}. It was shown that a large negative running of the spectral index of $\mathcal{O}(10^{-2})$ would make it difficult to support $50$ $e$--folds required for inflation \cite{Easther}. Designing inflationary models that predict a negative running with an acceptable $n_s$ and number of $e$--folds is not impossible. This occurs, for instance, in the inflationary potential with superimposed periodic oscillations \cite{Kobayashi}. Although by {\it Planck} data the central value of $\alpha_s$ is positive (if we include its derivative; \ie, $\beta_s$), this inflationary model has motivated us to study the possibility of PBHs formation in several inflationary models with an additional modulated term.  

The rest of the paper is organized as follows: In section 2, we first summarize the current bounds on the observational parameters and review their calculation from the potential of the inflaton field. In section 3, we briefly review the PBHs formation, and we find an upper bound on the running of the running of the spectral index. Then in section 4, we systematically analyze chaotic, inflection point, and hilltop inflationary models with an additional modulated term. Finally, the last section is devoted to discussion and conclusions.

Through this paper, we use the units $c=\hbar=\mpl=1$.

\section{Constraints on inflation parameters}

Within the slow--roll approximation any inflation model can be described by three independent parameters: the normalization of the curvature perturbation spectrum $\mathcal{P}_{\mathcal{R}_c}$, the tensor-to-scalar ratio $r$, and the spectral index $n_s$. If we wish to include higher-order effects, we have the forth and fifth parameters describing the running $\alpha_s$ and the running of running of the spectral index $\beta_s$, which has been measured for the first time by {\it Planck} \cite{Planck2}.

The power spectrum of the curvature perturbations is given by
\be\label{power1}
\mathcal{P}_{\mathcal{R}_{c}} = \dfrac{1}{12\pi^2 \mpl^6}\dfrac{V^3}{V^{\prime2}} \simeq  \dfrac{H^4}{4\pi^2 \dot{\phi}^2}\,,
\ee
where $\phi$ is the inflaton field and $V$ is its potential, and dot and prime denote the derivative with respect to the cosmic time and $\phi$, respectively. The power spectrum should be estimated at the time of horizon exit $k=aH$, and this normalization is quoted at the pivot scale $k_0 = 0.05\,$Mpc$^{-1}$, \footnote{This scale is roughly in the middle of the logarithmic range of scales probed by {\it Planck}. With this choice, $n_s$ is not strongly degenerate with the amplitude parameter $\mathcal{P}_{\mathcal{R}_c}$ \cite{Planck1}. Note that the ``COBE normalization scale'' is $k_{\rm COBE} = 0.002\,$Mpc$^{-1}$.} where $10^9\mathcal{P}_{\mathcal{R}_c}(k_0)=2.198\pm 0.056\,$ for {\it Planck}+WP\footnote{``WP'' means WMAP nine--years polarization low multipole likelihood at $\ell\leq 23$ \cite{WMAP9}.}+highL\footnote{``highL'' means the data from ACT \cite{ACT} and SPT \cite{SPT} in the range $\ell> 1000$.} data set \footnote{We use {\it Planck}+WP+highL data set because it has lower effective $\chi^2(=-2\Delta\ln(\mathcal{L}_{\rm max}))$ (see Table 5 of Ref.~\cite{Planck2}).}\cite{Planck2}.

On the other hand, the simplest assumption for the power spectrum is a scale-free power law
\be\label{power2}
\mathcal{P}_{\mathcal{R}_c}(k)=\mathcal{P}_{\mathcal{R}_c}(k_0)\left(\dfrac{k}{k_0}\right)^{n_s(k)-1}\,,
\ee
where $n_s(k)$ is the spectral index of the comoving curvature perturbation. However, if the initial perturbations are produced by inflation, then the power spectrum is not an exact powerlaw over all scales. Therefore, it is convenient to expand the spectral index $n_s$ as \cite{n_expansion} 
\be\label{n_expansion}
n_s(k)= n_s(k_0)+\dfrac{1}{2!}\,\alpha_s(k_0)\,\ln\lp\dfrac{k}{k_0}\rp+\dfrac{1}{3!}\,\beta_s(k_0)\,\ln^2\lp\dfrac{k}{k_0}\rp+\dots\,.
\ee
As we already mentioned, the parameters $\alpha_s$ and $\beta_s$ denote the running of the spectral index $n_s$ and the running of the running, respectively,
\bea
n_s(k_0)      &\equiv& \left.\dfrac{d\ln\mathcal{P}_{\mathcal{R}_c}}{d \ln k}\right|_{k=k_0}\,,\label{p1}\\
\alpha_s(k_0) &\equiv& \left.\dfrac{d n_s}{d\ln k}\right|_{k=k_0}\,,\label{p2}\\
\beta_s(k_0)  &\equiv& \left.\dfrac{d^2 n_s}{d\ln^2 k}\right|_{k=k_0}\,.\label{p3}
\eea
In the first order of slow--roll approximation, we can write the above parameters as follows \cite{n_expansion, running_of_running}:
\bea\label{nab}
n_s      &=& 1-6\epsilon+2\eta\,,\nn\\
\alpha_s &=& -24\epsilon^2+16\epsilon\eta-2\xi^2\,,\nn\\
\beta_s  &=& -192\epsilon^3+192\epsilon^2\eta-32\epsilon\eta^2-24\epsilon\xi^2+2\eta\xi^2+2\sigma^3\,,
\eea
where
\be\label{sl parameters}
\epsilon\equiv\dfrac{\mpl^2}{2}\lp\dfrac{V'}{V}\rp^{2}\,,\quad\eta\equiv\mpl^2\dfrac{V''}{V}\,,\quad\xi^2\equiv\mpl^4\dfrac{V' V'''}{V^2}\,,\quad\sigma^3\equiv\mpl^6\dfrac{V'^2V''''}{V^3}\nn
\ee
are the slow--roll parameters.

Another crucial parameter in inflationary models is the influence of gravitational waves relative to density perturbations, which is given by \cite{Lyth_book}
\be\label{tensor}
r\equiv\dfrac{C_2({\rm grav})}{C_2({\rm dens})}\simeq 16\epsilon\,.
\ee

The next step is to relate a value of $\phi$ to a comoving wave number $k$ that crosses the horizon at that value of $\phi$. To do so, we note that the number of $e$--foldings of inflation between a time $t$ and the end of inflation is
\be\label{efold1}
N(t) \equiv \ln\dfrac{a(t_{\rm end})}{a(t)} \simeq \dfrac{1}{\mpl^2}\int_{\phi_{\rm end}}^{\phi}\dfrac{V}{V'}d\phi\,,
\ee
which measures the amount of inflation that still has to occur after time $t$, with $N$ decreasing to $0$ at the end of inflation.
The expression for $N$ can be written as \cite{Planck2}
\be\label{efold2}
N\simeq 71.21-\ln\lp\dfrac{k}{a_0 H_0}\rp+\dfrac{1}{4}\ln\lp\dfrac{V_{\rm hor}}{\mpl^4}\rp+\dfrac{1}{4}\ln\lp\dfrac{V_{\rm hor}}{\rho_{\rm end}}\rp+\dfrac{1-3w_{\rm int}}{12(1+w_{\rm int})}\ln\lp\dfrac{\rho_{\rm reh}}{\rho_{\rm end}}\rp\,,
\ee 
where $\rho_{\rm end}$ is the energy density at the end of inflation, $\rho_{\rm reh}$ is a reheating energy scale after inflation, $a_0H_0$ is the present horizon scale, $V_{\rm hor}$ is the potential energy when the present horizon scale left the horizon during inflation, and $w_{\rm int}$ characterizes the equation of state between the end of inflation and the reheating energy scale. 
 
In the above equation, the first two terms are model independent, and if reheating happens rapidly, the magnitude of the last term is $\lsim\,1$. For most inflationary models, the third term $\sim-10$ and the fourth term is $\mathcal{O}(1)$ leading to the commonly assumed range $50 < N < 60$. We will consider the averaged value, therefore,
\be\label{efold3}
N(k) \simeq 55-\ln\lp\dfrac{k}{a_0H_0}\rp\,.
\ee

Now let us review the observational bounds on inflation parameters at the pivot scale of {\it Planck} data, $k_0=0.05$ Mpc$^{-1}$. With \p, a robust detection of the deviation ($\sim 6\sigma$) from scale invariance has been found with the base $\Lambda$CDM model \cite{Planck1}
\be\label{o1}
n_s = 0.959 \pm 0.007 \quad (68\,\%;\, \p+\rm{WP+highL})\,.
\ee

If we allow a possible running of the spectral index [the parameter $\alpha_s$ defined in eq.~\eqref{p2}], the {\it Planck} data give \cite{Planck2}
\bea\label{o2}
n_s      &=& 0.9548 \pm 0.0073  \quad (68\,\%;\, \p+\rm{WP+highL})\,,\nn\\
\alpha_s &=& -0.0149 \pm 0.0085 \quad (68\,\%;\, \p+\rm{WP+highL})\,.
\eea

The constraints are broadly similar if tensor fluctuations are allowed in addition to a running of the spectrum \cite{Planck2}:
\bea\label{o3}
n_s      &=& 0.9570 \pm 0.0075        \quad  (68\,\%;\, \p+\rm{WP+highL})\,,\nn\\
\alpha_s &=& -0.022^{+0.011}_{-0.010} \quad  (68\,\%;\, \p+\rm{WP+highL})\,,\nn\\
r &<& 0.23                            \quad  (68\,\%;\, \p+\rm{WP+highL})\,.
\eea

It is worth mentioning that by adding other cosmological data, the marginalized posterior distributions for $\alpha_s$ give \cite{Planck1}
\bea\label{o4}
\alpha_s=\left\{
\begin{array}{ll}
-0.013\pm 0.009\,     &(68\%;\,\p+{\rm{WP}})\,,\\
-0.011\pm 0.008\,     &(68\%;\,\p+{\rm{WP+highL+lensing}})
\end{array}
\right.
\eea
at the {\it Planck} pivot scale $k_0 = 0.05$\, Mpc$^{-1}$. Even with the tensor fluctuations in addition to a running of the spectrum, similar bounds are obtained \cite{Planck1}:
\bea\label{o5}
\alpha_s=\left\{
\begin{array}{ll}
-0.021\pm 0.011\,     &(68\%;\,\p+{\rm{WP}})\,,\\
-0.019\pm 0.010\,     &(68\%;\,\p+{\rm{WP+highL+lensing}})\,.
\end{array}
\right.
\eea

{\it Planck} also tested, for the first time, the possibility that the running depends on the wavelength so that $d\alpha_s/d\ln k$ is nonzero. With \p+WP+highL, the following results have been obtained at $68\,\%$ C.L.:
\bea\label{o6}
n_s      &=& 0.9476^{+0.086}_{-0.088}\,,\nn\\
\alpha_s &=& 0.001^{+0.013}_{-0.014}\,,\nn\\
\beta_s  &=& 0.022^{+0.016}_{-0.013}\,.
\eea

From eqs.\,\eqref{o2}-\eqref{o5} it is clear that when relaxing the assumption of a power-law spectrum, {\it Planck} data prefer a fairly large negative running for the scalar spectral index alone and in combination with other astrophysical data sets. But interestingly, when $\beta_s$ is allowed in the fit [\eqref{o6} data], the value for the running $\alpha_s$ is consistent with zero. If primordial, negative values for $dn_s/d\ln k$ of order $10^{-2}$ would be interesting for the physics of inflation. However, as already mentioned, a large negative running of the spectral index of $\mathcal{O}(10^{-2})$ would make it difficult to support the $N\approx 50\, e$--foldings required from inflation \cite{Easther}, but this holds only without nonzero derivatives higher than the third order in the inflationary potential. Designing inflationary models that predict a negative running with an acceptable $n_s$ and number of $e$--folds is not impossible. This occurs, for instance, in the axion monodromy model \cite{monodromy}, the inflaton potential with a steplike feature \cite{step_like} or multiple inflation \cite{multiple_inflation}. Another model which we will study in the next section is the inflationary potential with superimposed modulated oscillations \cite{Kobayashi}.

One of the important features of the {\it Planck} data set is that it hints at a significant ``running" in the running of the spectral index, and as studied in detail in Refs.~\cite{Encieh1, Encieh2}, this parameter is crucial for DM PBHs formation.

In the following section, we briefly review the PBHs formation in inflationary models and find that the nonproduction of (long--lived) PBHs puts a stronger upper bound on $\beta_s$.

\section{Primordial black holes}

PBHs are black holes that could be formed by many different mechanisms during the very early universe, \eg, from initial density inhomogeneities \cite{Zeldovich}, a softening of the equation of state \cite{khlopov}, collapse of cosmic string loops \cite{Hawking2}, bubble collisions \cite{bubble}, collapse of domain walls \cite{domain walls}, etc.
They are an important tool in cosmology to probe the primordial spectrum of small scale curvature perturbations that reenter the cosmological horizon during the radiation-dominated era after inflation. On the other hand, the PBHs with mass larger than $10^{15}$ g, which survive the Hawking radiation \cite{Hawking} till present time, could be candidates for dark matter.

The traditional treatment of PBH formation is based on the Press-Schechter formalism \cite{Press-Schechter} used widely in large scale structure studies. In Ref.~\cite{Encieh1}, this mechanism has been studied in detail for the formation of PBHs in the radiation-dominated era just after inflation. There, we found that for the formation of PBHs [with mass larger than $10^{15}$ g, which could form (part of) the cold dark matter in the Universe] from density perturbations with a power-law spectrum, the spectral index at scale of PBHs formation ($k_{\rm PBH}$)\footnote{The relevant scale for the formation of $10^{15}$ g PBHs is $k_{\rm PBH}\simeq 10^{15}$ Mpc$^{-1}$ \cite{Encieh1}.} should be at least $1.37$. \footnote{It is worth mentioning that this value is completely independent of any inflationary model, and the only assumption is that the density perturbations are originated from the slow--roll phase of inflation.} Therefore, DM PBHs formation can only happen if the spectral index increases significantly between the scales probed by CMB (or other cosmological observations) and the relevant scales for DM PBHs. With $n_s(k_{\rm PBH})\simeq 1.37$, the power spectrum at the scale of DM PBHs formation should be $\mathcal{P}_{\mathcal{R}_c}(k_{\rm PBH})\simeq 10^{-3}$, which is $\sim10^6$ times higher than the value of the power spectrum in the observed CMB scales. 
According to the observational data, the spectral index at the pivot scale has red tilt and its first derivative $\alpha_s$ is negative; therefore, for PBHs formation, the positive value of the second derivative of the spectral index $\beta_s$ at the pivot scale of the observational data is required. However, this is a necessary but not a sufficient condition, since for PBHs formation, the power spectrum should increase (\ie, it should reach $\sim10^{-3}$) at scales relevant for DM PBHs formation.
If we consider the observational results where the running of the running of the spectral index is not considered in the fit, \ie, eq.~\eqref{o2} data and we set $\alpha_s$ equal to its central value $\alpha_s(k_0) = -0.0149$, eq.~\eqref{n_expansion} shows that in order not to generate large density perturbations for DM PBHs formation, we only need
\be\label{bound1}
\beta_s(k_{0}) \lsim 0.0025
\ee
Even if we consider the eq.~\eqref{o6} data, we only need
\be\label{bound2}
\beta_s(k_{0}) \lsim 0.0017\,.
\ee
These results are one order of magnitude smaller than the central value of $\beta_s$ reported by \p+WP+highL data set. Hence, the nonproduction of DM PBHs puts a stronger upper bound on the running of the running of the spectral index.

In the next section, after explaining the idea of inflationary model with superimposed modulated term, we study several models in this concept and check whether these models can accommodate the values of inflation parameters (\ie, $\mathcal{P}_{\mathcal{R}_c},\,n_s,\,\alpha_s$, and $\beta_s$), as indicated by current data. As a by--product, we also check whether they can lead to long--lived PBHs formation. 

\section{Inflation Models}

In this section, we first explain the idea of inflation potential with an additional superimposed periodic term, and then we study the (linear and quadratic) chaotic inflaton potentials, inflection point, and hilltop inflationary models in this concept. We focus on these models because they have been studied in Ref.~\cite{Encieh2} after WMAP seven-years data \cite{WMAP-7} in the case of PBHs formation. We recheck whether these models with a new additional modulated term can lead to the formation of DM PBHs after the recent CMB data. It is worth mentioning that some of these models are not consistent with {\it Planck} data (for a review on preferred inflationay models after {\it Planck} data, see Ref.~\cite{models after Planck}). However, as we will show, when a modulated term is added, these models will be consistent with the recent data.

\subsection{Idea}

We consider an inflaton potential $V(\phi)$ with an additional modulated term, and we decompose the potential as \cite{Kobayashi}
\be\label{potential}
V(\phi)=V_0(\phi)+V_{\rm mod}(\phi)\,,
\ee
where the first term is a smooth potential and the second term represents the modulation. An important assumption is that the overall inflation dynamics (such as the Hubble parameter and the inflaton velocity) should not be altered by the modulation; this means that the effect of the modulation $V_{\rm mod}$ (and its derivatives) on the inflaton dynamics is sufficiently small when averaged over a sufficiently long time or large field space, \ie,
\bea\label{c1}
\left|V_0(\phi)\right| &\gg& \left|V_{\rm mod}(\phi)\right|\,,\\
\left|V'_0(\phi)\right| &>& \left|V'_{\rm mod}(\phi)\right|\,.
\eea
We also suppose that the slow--roll approximations
\be\label{sra}
3H\,\dot{\phi} \simeq -V^{\prime}(\phi)\, , \qquad 3H^2 \simeq V(\phi)
\ee
are valid (where $H$ is the Hubble parameter). In order for the running (and the running of the running) of the spectral index to be large enough such that the spectral tilt (and its running) switches between red and blue within the observed CMB scales, we further require
\bea\label{c2}
\left|V''_0(\phi)\right| &\lsim& \left|V''_{\rm mod}(\phi)\right|\,,\\
\left|V'''_0(\phi)\right| &\ll& \left|V'''_{\rm mod}(\phi)\right|\,,\\
\left|V''''_0(\phi)\right| &\ll& \left|V''''_{\rm mod}(\phi)\right|\,.
\eea
To be explicit, we are considering a hierarchy
\be\label{hierarchy}
\left|\dfrac{V_{\rm mod}}{V_0}\right| \ll \left|\dfrac{V'_{\rm mod}}{V'_0}\right| \ll \left|\dfrac{V''_{\rm mod}}{V''_0}\right| \ll \left|\dfrac{V'''_{\rm mod}}{V'''_0}\right| \ll \left|\dfrac{V''''_{\rm mod}}{V''''_0}\right|\,.
\ee
Note that these conditions are only required during inflation and need not hold along the whole inflaton potential.

An important assumption here is that the amplitude of the modulation term $V_{\rm mod}(\phi)$ on the inflaton dynamics is sufficiently small when averaged over a sufficiently long time or large field space; therefore, the inflaton does not get stuck in one of the local minima introduced in the potential by the oscillations. One example for such modulations is a sine function, and we will see that the above conditions can be satisfied in this case.

In the presence of the modulated term, the spectral index, its running, and its running of running are easily modified by the modulation, while not affecting the overall inflaton dynamics and can be estimated as follows:
\bea\label{srp}
n_s      &\simeq& 1-6\epsilon_0+2\eta_0-\mpl^2\bigg(6\dfrac{V'V'_{\rm mod}}{V_0^2}+2\dfrac{V''_{\rm mod}}{V_0}\bigg)\,,\label{par.a}\\
\alpha_s &\simeq& -24\epsilon_0^2+16\epsilon_0\eta_0-2\xi_0^2-\mpl^4\bigg(24\dfrac{V'^3V'_{\rm mod}}{V_0^4}+8\dfrac{V'^2V''_{\rm mod}}{V_0^3}-2\dfrac{V'V'''_{\rm mod}}{V_0^2}\bigg)\,,\label{par.b}\\
\beta_s  &\simeq& -192\epsilon_0^3+192\epsilon_0^2\eta_0-32\epsilon_0\eta_0^2-24\epsilon_0\xi_0^2+2\eta_0\xi_o^2+2\sigma_0^3-\mpl^6\bigg(24\dfrac{V'^5V'_{\rm mod}}{V_0^6}\label{par.c}\\
&+&48\dfrac{V'^{3}V'_{\rm mod}V''_{\rm mod}}{V_0^5}-16\dfrac{V'V'_{\rm mod}V''^2_{\rm mod}}{V_0^4}-12\dfrac{V'^2V'_{\rm mod}V'''_{\rm mod}}{V_0^4}+2\dfrac{V'V''_{\rm mod}V'''_{\rm mod}}{V_0^3}+2\dfrac{V'^2V''''_{\rm mod}}{V_0^3}\bigg)\,,\nn
\eea
where slow--roll parameters with the subscript ``0'' are contributions from the smooth potential $V_0$. Note that in the above expressions, we have kept contributions from the smooth potential and also the leading oscillatory terms due to the modulation $V_{\rm mod}$ by using the hierarchy \eqref{hierarchy}; \ie, in eq.~\eqref{par.a}, the $\eta_0$ and $V''_{\rm mod}/V_0$ terms are dominant, in eq.~\eqref{par.b}, the $2\xi_0^2$ and $(V'V'''_{\rm mod})/V_0^2$ terms are dominant, and in $\beta_s$ [eq.~\eqref{par.c}], the $2\eta_0\xi_o^2,\,2\sigma_0^3$ and $(V'^2V''''_{\rm mod})/V_0^3$ terms play the significant role. 

In the case of the power spectrum of curvature perturbations \eqref{power1}, since $\dot{\phi}$ is determined by $V'$ [see eq.~\eqref{sra}], the main contribution also comes from the $V'_0$, and the $V'_{\rm mod}$ causes only small oscillations on the power spectrum, as we will see below for several inflationary models. 

In the following, we consider explicit inflationary models with superimposed oscillatory term, and we check the consistency of these models with recent observational data. We also study the possibility of (long-lived) PBHs formation in these models.

\subsection{Power-law (Chaotic) Inflation}

The simplest class of inflationary models is characterized by a power-law potential of the form \cite{Linde1983}
\be\label{model11}
V_0(\phi)=\rho^4\lp\dfrac{\phi}{\mu}\rp^p\,,
\ee
where $\rho$ and $\mu$ are constants, and $p$ is an integer. This class of potentials includes the simplest chaotic models, in which inflation starts from large values for the inflaton, $\phi > \mpl$. Inflation ends by violation of the slow--roll regime, and we assume this occurs at $\epsilon=1$. A linear potential with $p=1$ \cite{monodromy}\footnote{This model is known as the axion monodromy inflation model \cite{monodromy}.} has $\eta=0$ and lies within the $95\%$ C.L. region in the $n_s-r$ plane for the \p+WP+highL data \cite{Planck2}. Inflation with $p=2/3$ \cite{Silverstein} lies on the boundary of the joint $95\%$ C.L. region. The model with a quadratic potential $p=2$ \cite{Linde1983} lies outside the joint $95\%$ C.L. for $N \lsim 60$ $e$--folds. The $p=4$ model lies well outside of the joint $99.7\%$ C.L. region in the $n_s-r$ plane after the {\it Planck} data \cite{Planck2}.

With superimposed periodic oscillation, we rewrite the potential as follows:
\be\label{model12}
V(\phi)=\lambda\phi^p+\Lambda^4\cos\lp\dfrac{\phi}{f}+\theta\rp\,,
\ee
where $\Lambda,\,f$ and $\theta$ are constants and $\lambda\equiv\rho^4/\mu^p$. We assume that the period of oscillation $f$ is sub-Planckian, \ie, $f<1$. We also require that one oscillation period corresponds to the time duration when the observed scales exited the horizon.

Assuming that the slow--roll approximations \eqref{sra} are valid, the inflation parameters can be estimated as follows\footnote{In Ref.~\cite{Kobayashi}, this model has been studied without a $\beta_s$ parameter to accommodate the WMAP seven-years data \cite{WMAP7}.}:
\bea\label{model13}
n_s      &\simeq& 1-\lp\dfrac{\mpl}{\phi}\rp^2 \left[p(p+2)+\dfrac{2\Lambda^4}{\lambda f^2\phi^{p-2}}\cos\lp\dfrac{\phi}{f}+\theta\rp\right]\,,\nn\\
\alpha_s &\simeq& -2p\lp\dfrac{\mpl}{\phi}\rp^4 \left[p(p+2)+\dfrac{\Lambda^4}{\lambda f^3\phi^{p-3}}\sin\lp\dfrac{\phi}{f}+\theta\rp\right]\,,\nn\\
\beta_s  &\simeq& -2p^2\lp\dfrac{\mpl}{\phi}\rp^6 \left[4p(p+2)-\dfrac{\Lambda^4}{\lambda f^4\phi^{p-4}}\cos\lp\dfrac{\phi}{f}+\theta\rp\right]\,,
\eea
and the spectrum of the curvature perturbation is given by
\be\label{model14}
\mathcal{P}_{\mathcal{R}_c}\simeq \dfrac{\lambda}{12\pi^2 p^2 \mpl^6}\phi^{p+2}\left[1-\dfrac{\Lambda^4}{p\lambda f\phi^{p-1}}\sin\lp\dfrac{\phi}{f}+\theta\rp\right]^{-2}\,,
\ee 
which should be estimated when the scales leave the horizon. In this model, we also assume that $\Lambda^4/(p\lambda\,f\phi^{p-1})<1$, which guarantees that during inflation the inflaton field is not trapped by the modulations.

In the final expressions \eqref{model13} and \eqref{model14}, we have only kept the leading oscillatory terms. It is clear that the spectral index, its running, and its running of running oscillate with growing oscillation amplitudes as inflation proceeds and the inflaton field becomes smaller. Note also that the power of $f$ appearing in eq.~\eqref{model13} are larger than that of eq.~\eqref{model14}, and that is why the spectral index and its derivatives can be significantly 
modified without affecting the power spectrum too much.

Now we set the parameters of the potential such that the observable parameters satisfy the bounds from \p+WP+highL data as reported in eq.~\eqref{o6}, as well as the value of the power spectrum at the pivot scale of \p. We study the linear $p=1$ and the quadratic $p=2$ potentials. The results for the parameters are in Table~\ref{tab:1}, where $\phi_{_0}$ is the value of the inflaton field when leaves the {\it Planck} pivot scale. 
\begin{table}[tbp]
\centering
\begin{tabular}{ | c | c | c | }
  \cline{2-3}
  \multicolumn{1}{c|}{}    & $p=1$                    & $p=2$                 \\\cline{1-3}
  $\lambda$                & $2.9\times10^{-10}$      & $2.3\times10^{-11}$   \\\cline{1-3}
  $\Lambda$                & $1.79\times10^{-3}$      & $2.0\times10^{-3}$    \\\cline{1-3}
	$\phi_{_0}$              & $10.02$                  & $14.28$               \\\cline{1-3}
	$f$                      & $0.242$                  & $0.340$               \\\cline{1-3}
	$\theta$                 & $2.32$                   & $2.44$                \\\cline{1-3}
\end{tabular}
\caption{Parameter values of the linear ($p=1$) and the quadratic ($p=2$) potentials.}
\label{tab:1}
\end{table}

\begin{figure}[h!]
\centering{\includegraphics[width=0.45\textwidth]{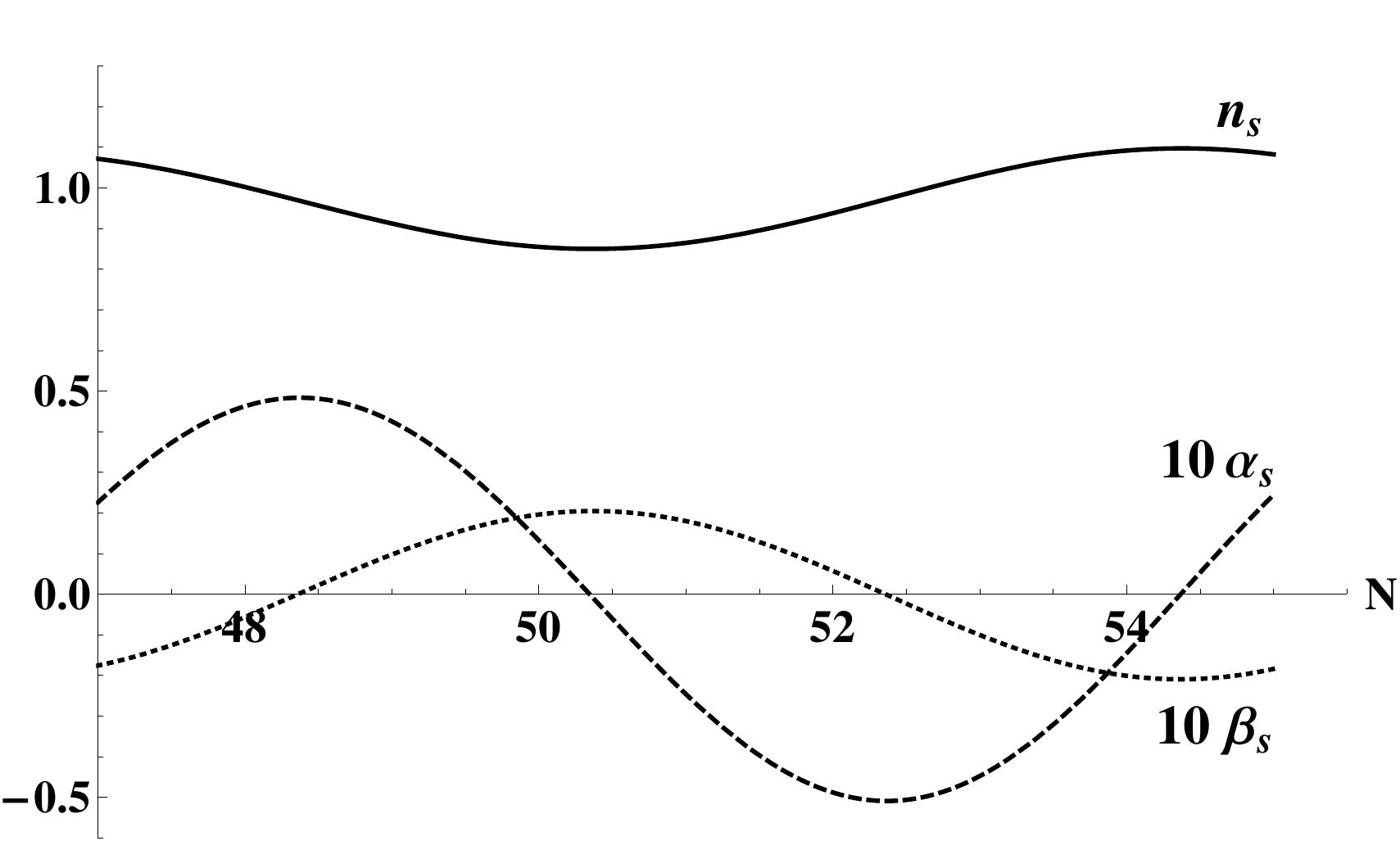}
\hspace{0.3cm}
\includegraphics[width=0.45\textwidth]{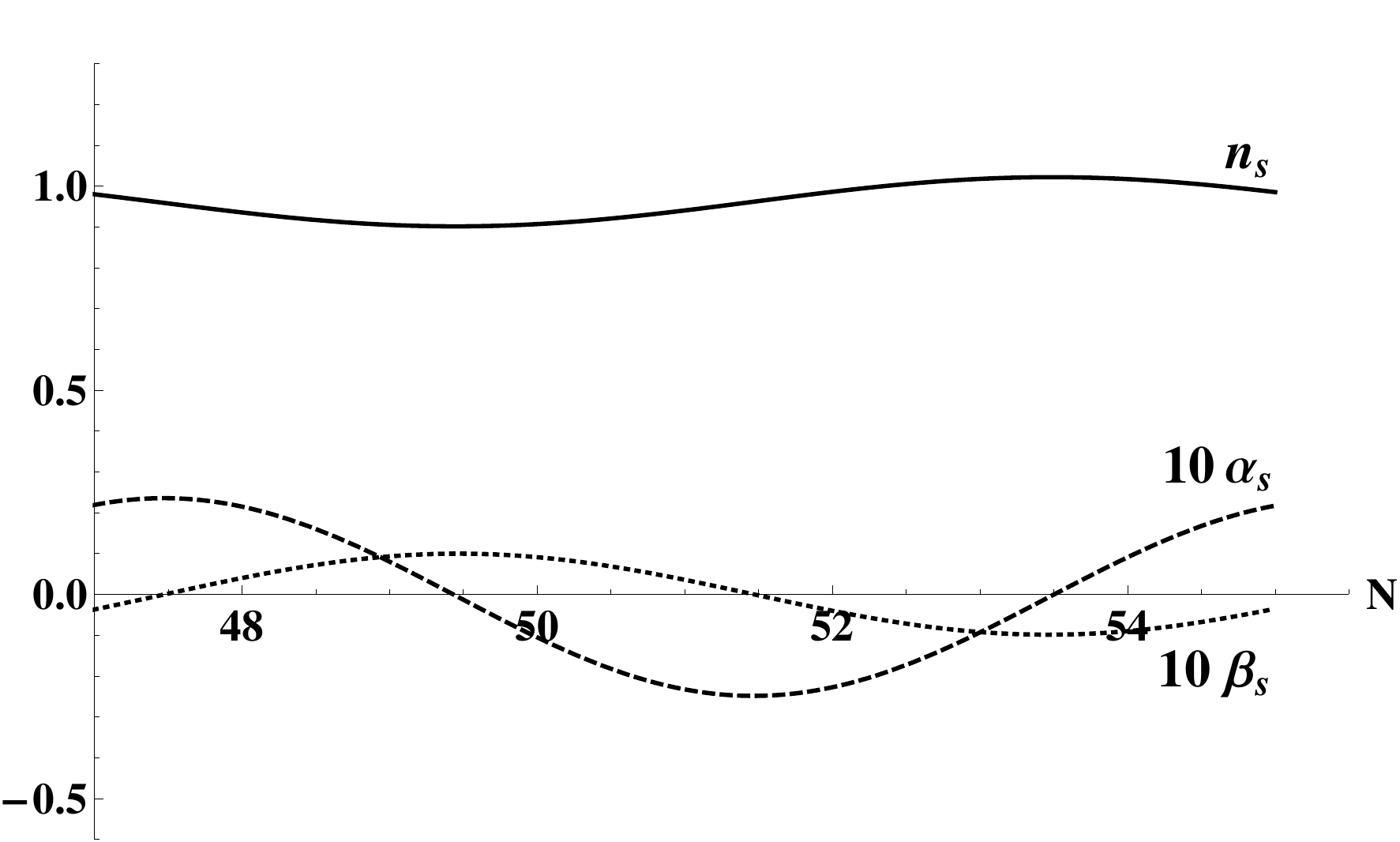}}
\caption{The left (right) figure shows the inflation parameters [\ie, $n_s$ (solid line), $\alpha_s$ (dashed line), and $\beta_s$ (dotted line)] with respect to the number of $e$--folds in linear (quadratic) potential.}
\label{fig:par11}
\end{figure}

In Figure~\ref{fig:par11}, we show the inflation parameters (\ie, $n_s,\,\alpha_s$, and $\beta_s$) with respect to the number of $e$--folds - when observable scales leave the horizon - in linear (left plot) and quadratic (right plot) potentials. Note that the observable scales correspond to about eight $e$--folds of inflation, \ie, $47 \leq N \leq 55$. It is clear that due to the modulated term, these parameters have oscillatory behavior, and as we mentioned before, one period of their oscillation corresponds to the time duration (\ie, number of $e$--folds) when the observed scales leave the horizon.

\begin{figure*}[h!]
\includegraphics[width=0.45\textwidth]{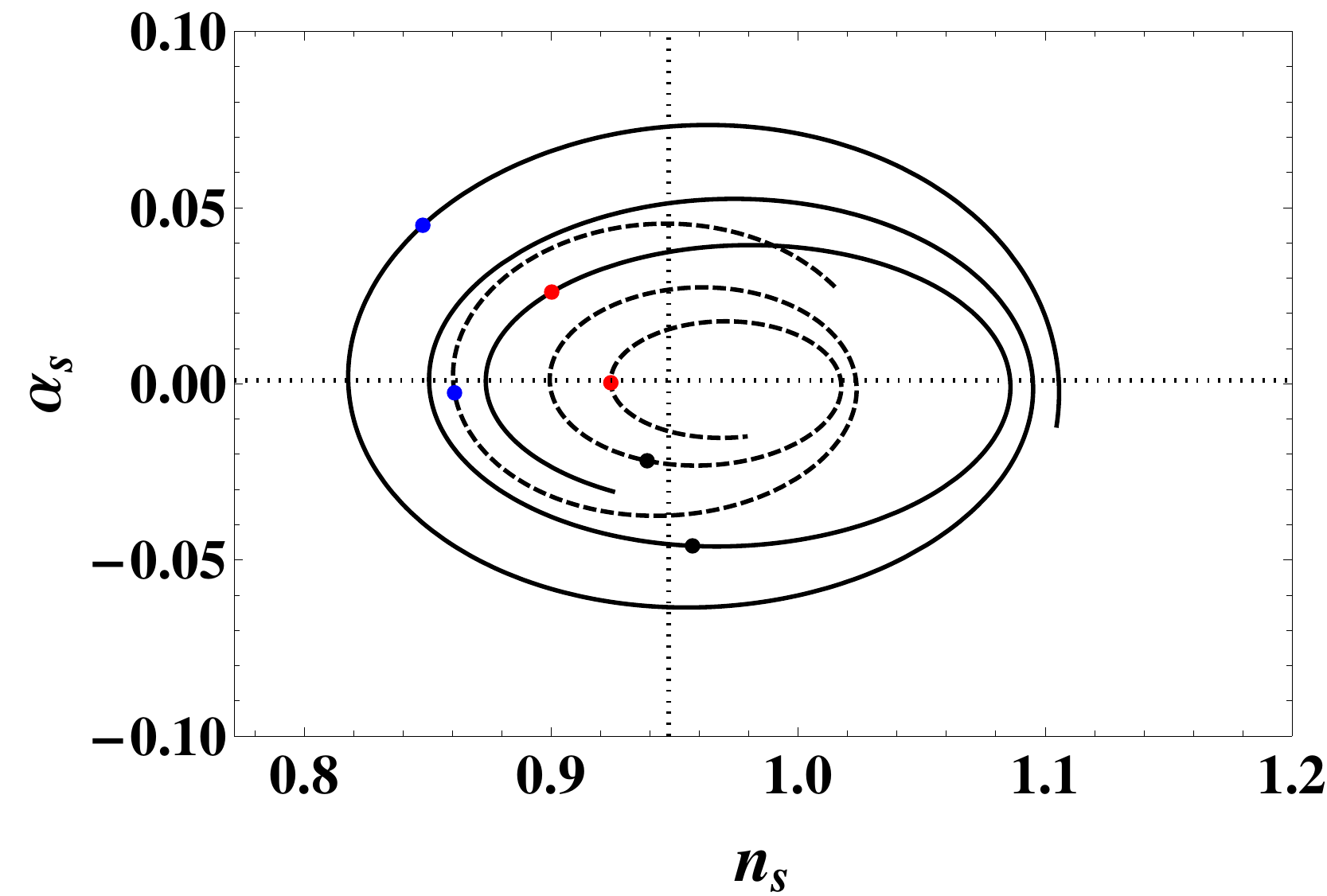}
\vspace{-5.9cm}
\begin{center}
\hspace*{6.2cm}
\includegraphics[width=.58\textwidth]{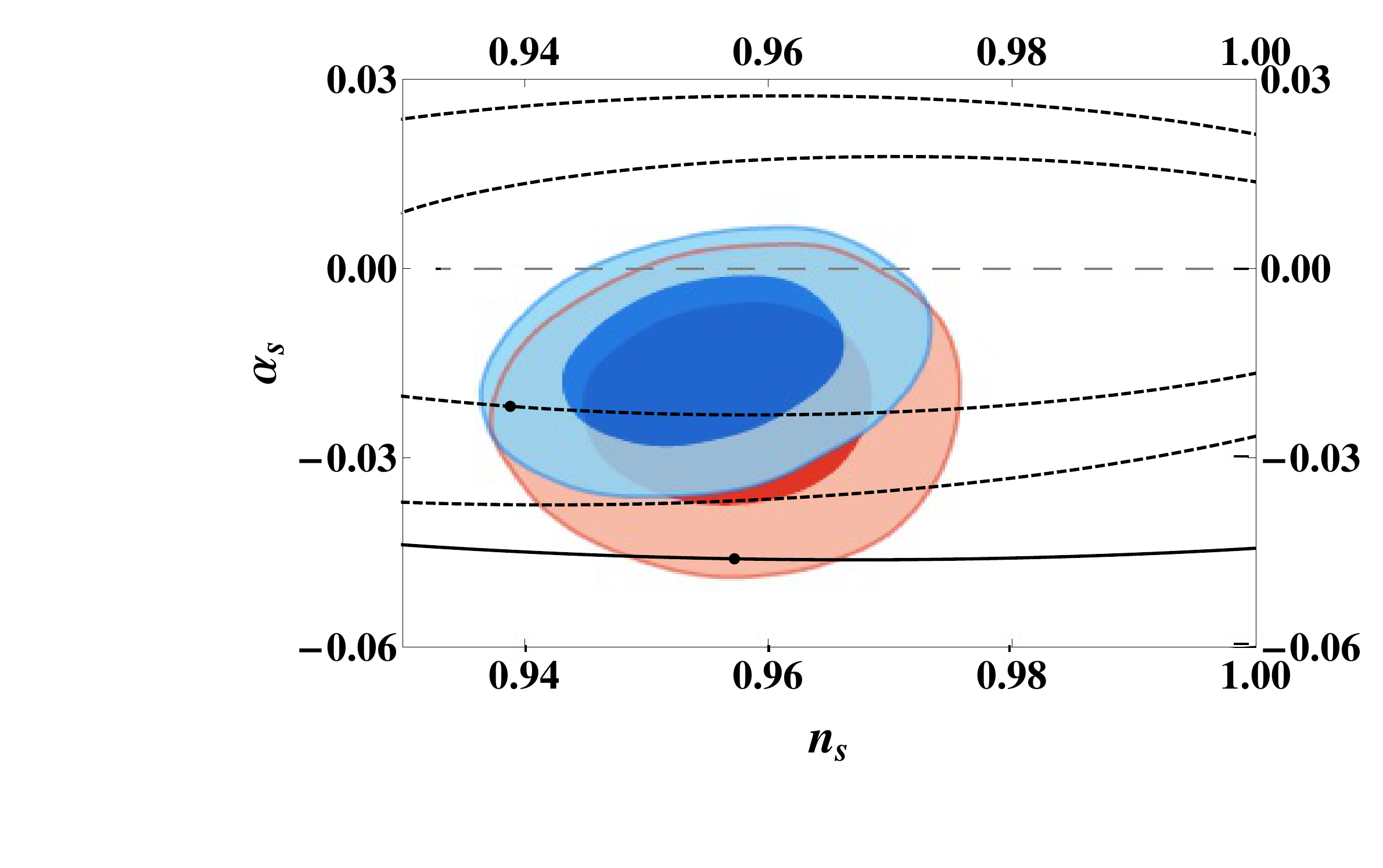}
\end{center}
\vspace{-1.cm}
\caption{Variation of the spectral index $n_s$ and its running $\alpha_s$ as inflaton proceeds along the linear (solid curve) and the quadratic (dashed curve) potentials with modulation. In the left figure the dotted lines are the central values of the parameters as reported in eq.~\eqref{o6}, and the curves start from 70 $e$--folds before the end of inflation (end point of innermost orbits) up to 30 $e$--folds before the end of inflation. Black points correspond to the pivot scale (about 50 $e$--folds before the end of inflation). The scale which exited the horizon 15 $e$--folds before (after) the pivot scale is indicated by red (blue) points. In the right figure, the superimposed contours are ($68\%$ and $95\%$ C.L.) constraints from \p+WP+highL data \cite{Planck1} with running (blue) and additionally with tensors (red).}
\label{fig:par12}
\end{figure*}

\begin{figure}[h!]
\hspace*{-2.cm}
\vspace*{-1.5cm}
\includegraphics[width=1.\textwidth]{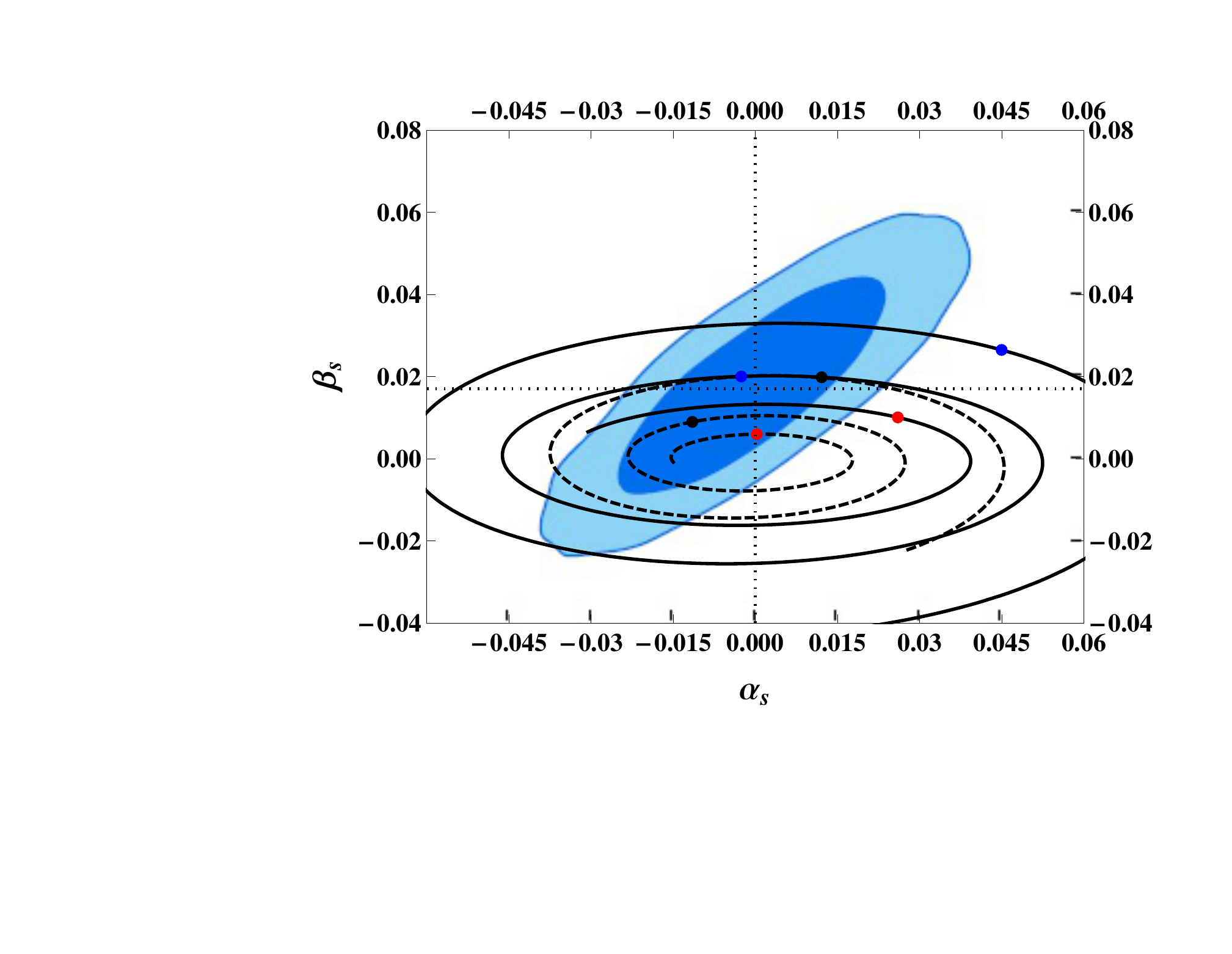}
\vspace*{-1.25cm}
\caption{Variation of the running of the spectral index $\alpha_s$ and its running $\beta_s$ as inflaton proceeds along the linear (solid curve) and quadratic (dashed curve) potentials with modulation. The dotted lines are the central values of the parameters as reported in \p+WP+BAO data, and the curves start from 70 $e$--folds before the end of inflation (end point of innermost orbits) up to 30 $e$--folds before the end of inflation. Black points correspond to the pivot scale (about 50 $e$--folds before the end of inflation). The scale which exited the horizon 15 $e$--folds before (after) the pivot scale is indicated by red (blue) points. The superimposed contour is marginalized joint $68\%$ and $95\%$ C.L. regions using \p+WP+BAO data \cite{Planck2}.}
\label{fig:par13}
\end{figure}

In Figure~\ref{fig:par12}, we plot oscillating trajectories in the $n_s-\alpha_s$ plane for the linear $(p=1)$ and the quadratic $(p=2)$ potentials with modulation and the varying running and its running (\ie, $\alpha_s-\beta_s$ plane) are shown in Figure~\ref{fig:par13}.\footnote{In the {\it Planck} paper, \cite{Planck2} the marginalized joint $68\%$ and $95\%$ C.L. regions in the $\alpha_s-\beta_s$ plane have shown only for \p+WP+{\bf BAO} data. According to Table~5 of Ref.~\cite{Planck2}, the inflation parameters in this data set are as follows:
\begin{eqnarray*}\label{o7}
n_s      &=& 0.9568^{+0.068}_{-0.063}\,,\\\nn
\alpha_s &=& 0.000^{+0.063}_{-0.016}\,,\\\nn
\beta_s  &=& 0.017^{+0.016}_{-0.014}\,.\nn
\end{eqnarray*} 
It is clear that the central values of these data are similar to the \p+WP+{\bf highL} data [see eq.~\eqref{o6}].} Note that the marginalized joint $68\%$ and $95\%$ C.L. regions in the $n_s-\alpha_s$ plane (right plot of Figure~\ref{fig:par12}) are without considering the $\beta_s$ parameter in data, \ie, $n_s+\alpha_s$ model without tensor modes, eq.~\eqref{o2} data (blue contours), and $n_s+\alpha_s$ model with tensor modes, eq.~\eqref{o3} data (red contours). 
 
From Figure~\ref{fig:par13}, it is clear that the value of the $\beta_s$ at the pivot scale (black points in Figure~\ref{fig:par13}) is positive; therefore, we study the possibility of DM PBHs in these models.
With the parameter sets in Table~\ref{tab:1} and eq.~\eqref{n_expansion}, one can easily calculate the value of the spectral index at the scale of DM PBHs formation in the chaotic inflationary model. The results are as follows:
\bea
p=1 \quad &\Rightarrow& n_s(k_{\rm PBH})\simeq 5.85 > 1.37\,,\nn\\
p=2 \quad &\Rightarrow& n_s(k_{\rm PBH})\simeq 2.84 > 1.37\,.\nn
\eea

It is clear that in the $p=1$ and $p=2$ chaotic models, the value of the spectral index at the scale of DM PBHs is larger than $1.37$; therefore, PBHs can form in these models.

The above results can also be seen by directly computing the power spectrum at the scale of ehe DM PBHs formation.
With the parameter sets in Table~\ref{tab:1}, we numerically solve the equation of motion of the inflaton field and the Friedmann equation [eq.~\eqref{sra}] for linear and quadratic potentials and compute the power spectrum \eqref{power1}, which are shown in figures~\ref{fig:chaotic1} and \ref{fig:chaotic2}, respectively. In the following figures, the blue solid curves and black solid lines are the power spectrum with and without the modulated term, respectively. For comparison, we also plotted the power spectrum by including the running $\alpha_s$ (dashed blue curves) and the running of the running $\beta_s$ (dotted blue curves). One can clearly see the oscillatory features in the power spectrum.

Note that the \p+WP+highL observations probe scales from $k_{H_0} \sim 2.22\times 10^{-4}$ Mpc$^{-1}$ down to $\sim 1.12$ Mpc$^{-1}$, and from the left plots, it is clear that within this range these curves are almost degenerate. The difference become evident only at small scales $k\gsim10$ Mpc$^{-1}$. To clarify this point further, we also plotted the power spectrum in a wider range [\ie, the range of the CMB observation and DM PBHs formation ($\sim10^{15}$ Mpc$^{-1}$)]. From the right plots, it is clear that if we only consider the running of the spectral index (dashed blue curves), the power spectrum can not reach $\sim10^{-3}$, which is the relevant value for DM PBHs formation. However, when the running of the running is included (dotted blue curves), the power spectrum can increase significantly at small scales (\ie, large wave numbers), and in $p=1$ and $p=2$, chaotic models can reach the required value for DM PBHs formation.

\begin{figure}[h!]
\centering{\includegraphics[width=0.48\textwidth]{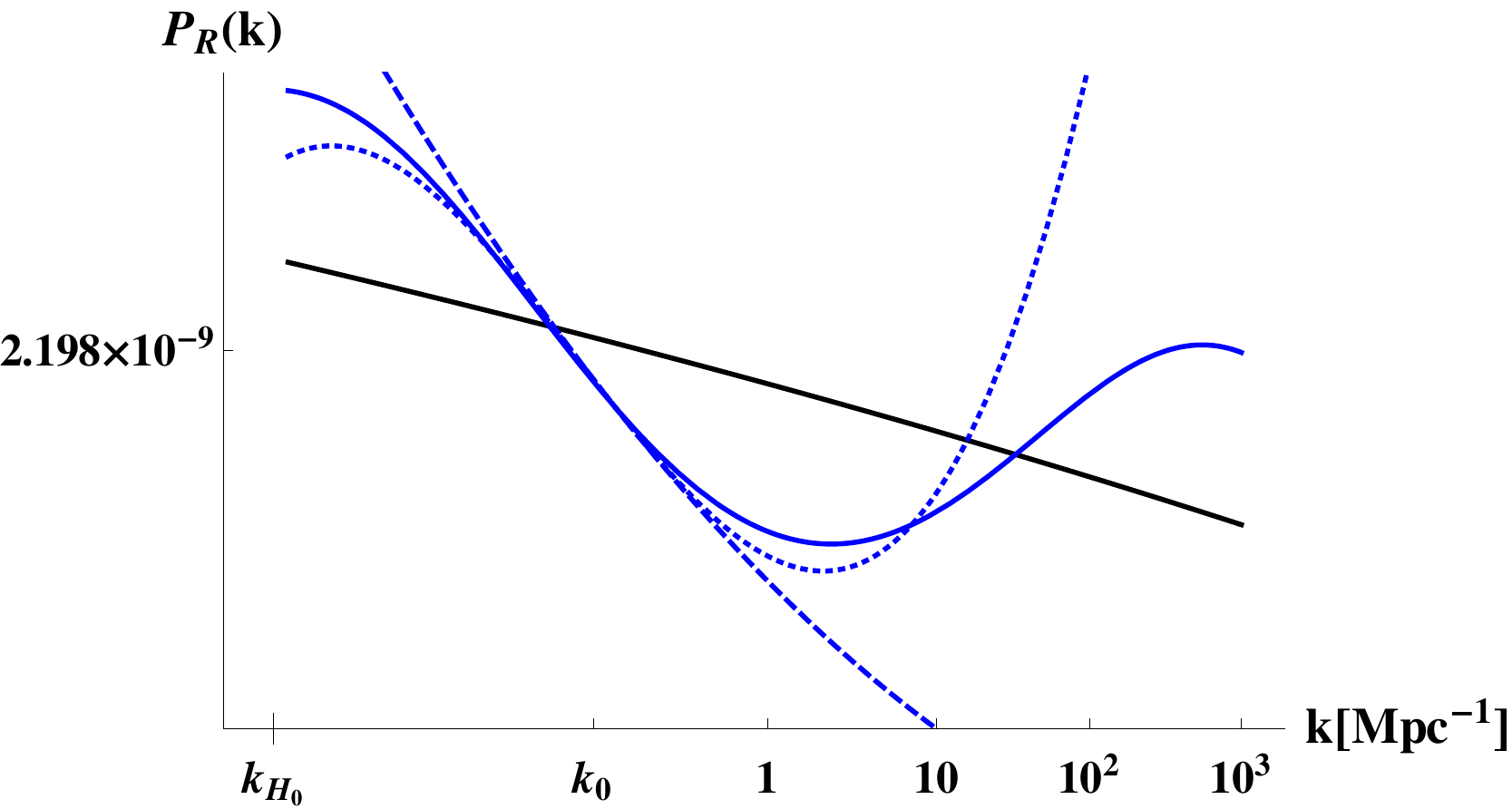}
\hspace{0.1cm}
\includegraphics[width=0.48\textwidth]{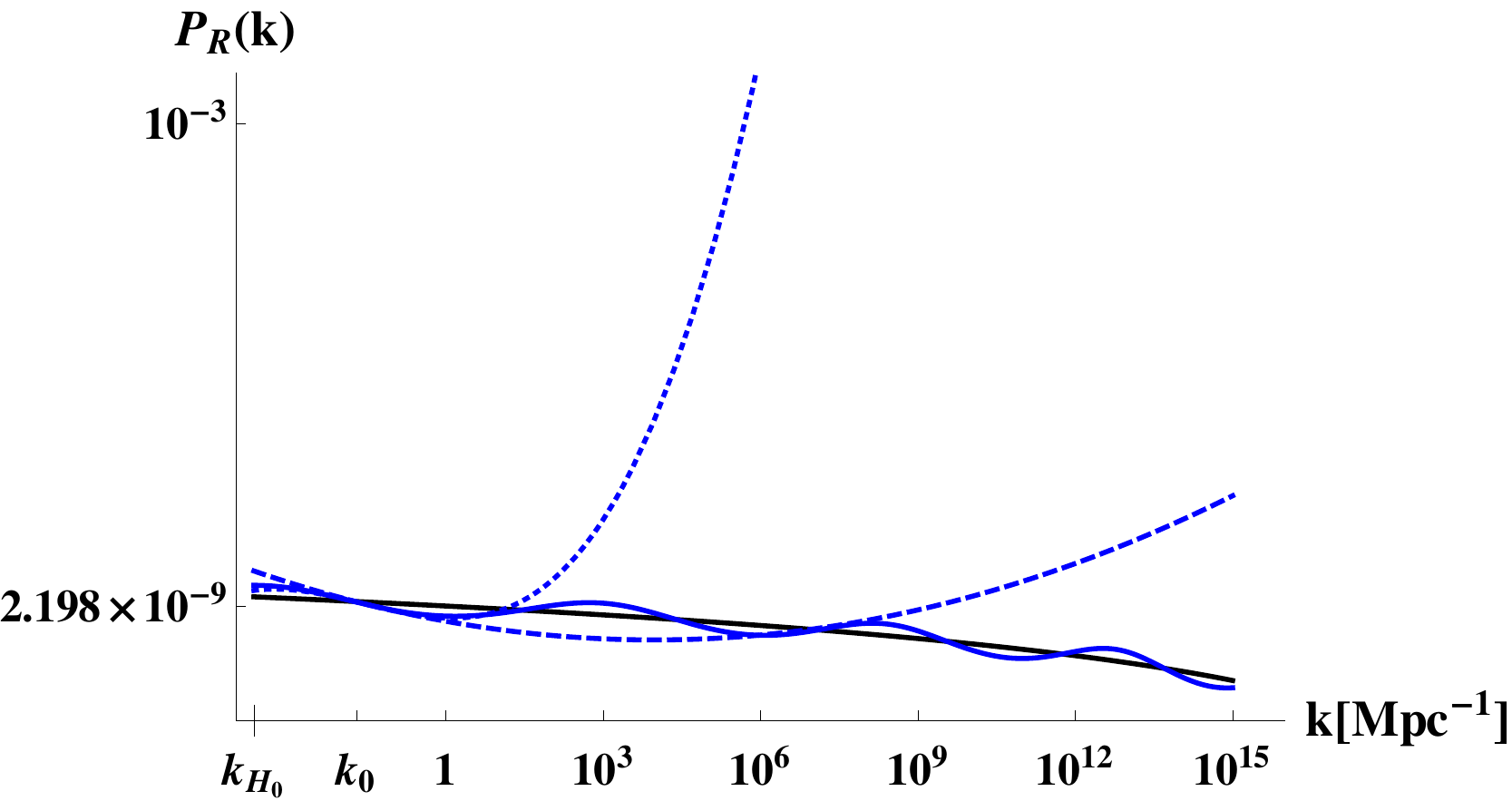}}
\caption{Curvature perturbation spectrum from the inflaton potential \eqref{model12} for $p=1$ (solid blue curve). Spectrum by including the running (of the running) is shown as the blue dashed (dotted) curve. Spectrum from a potential without any modulation is shown as the black line for comparison. Note that all the curves realize the observed value of the spectral index at the pivot scale $k_{0}=0.05$ Mpc$^{-1}$.}
\label{fig:chaotic1}
\end{figure}

\begin{figure}[h!]
\centering{\includegraphics[width=0.48\textwidth]{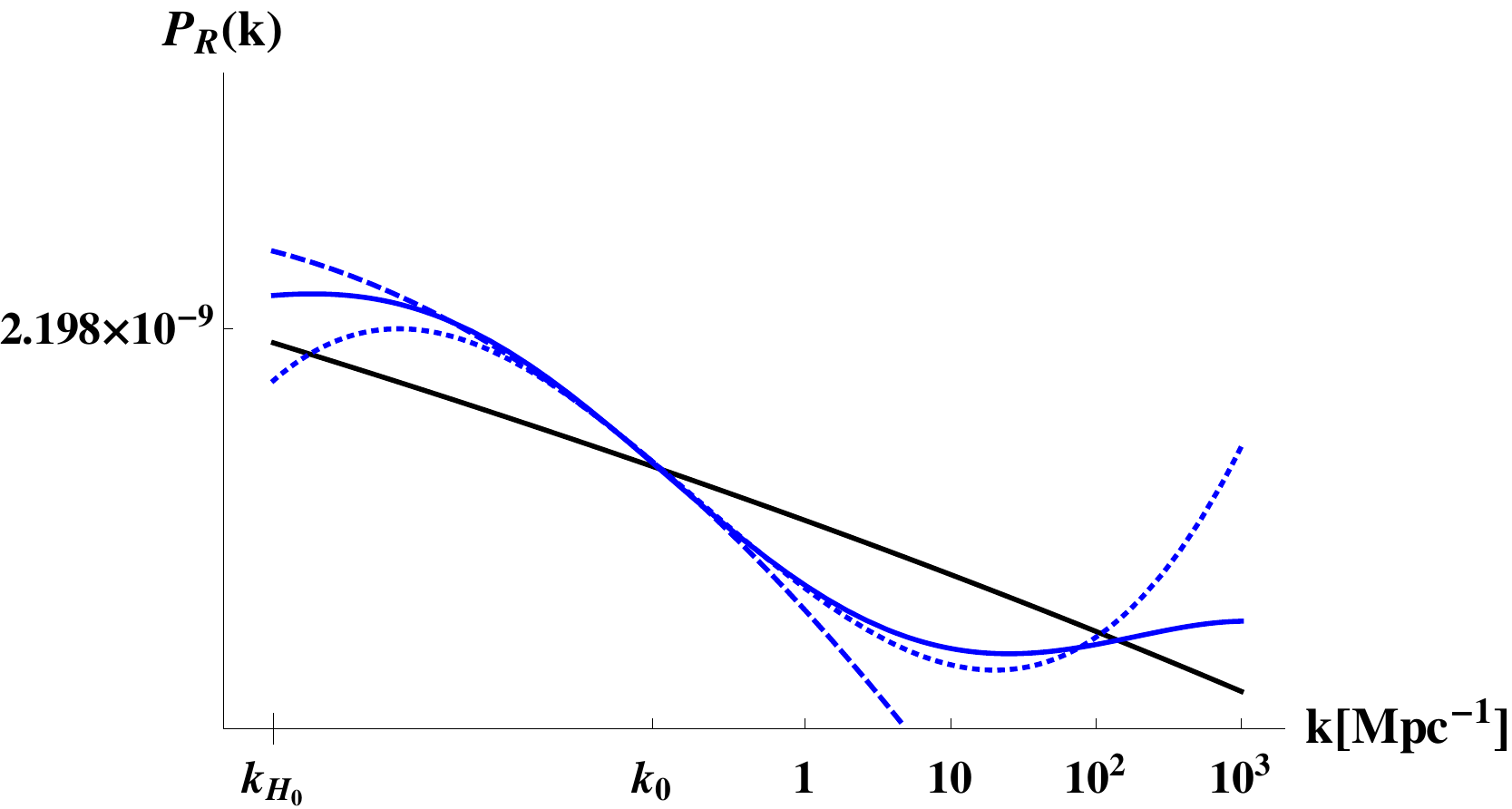}
\hspace{0.1cm}
\includegraphics[width=0.48\textwidth]{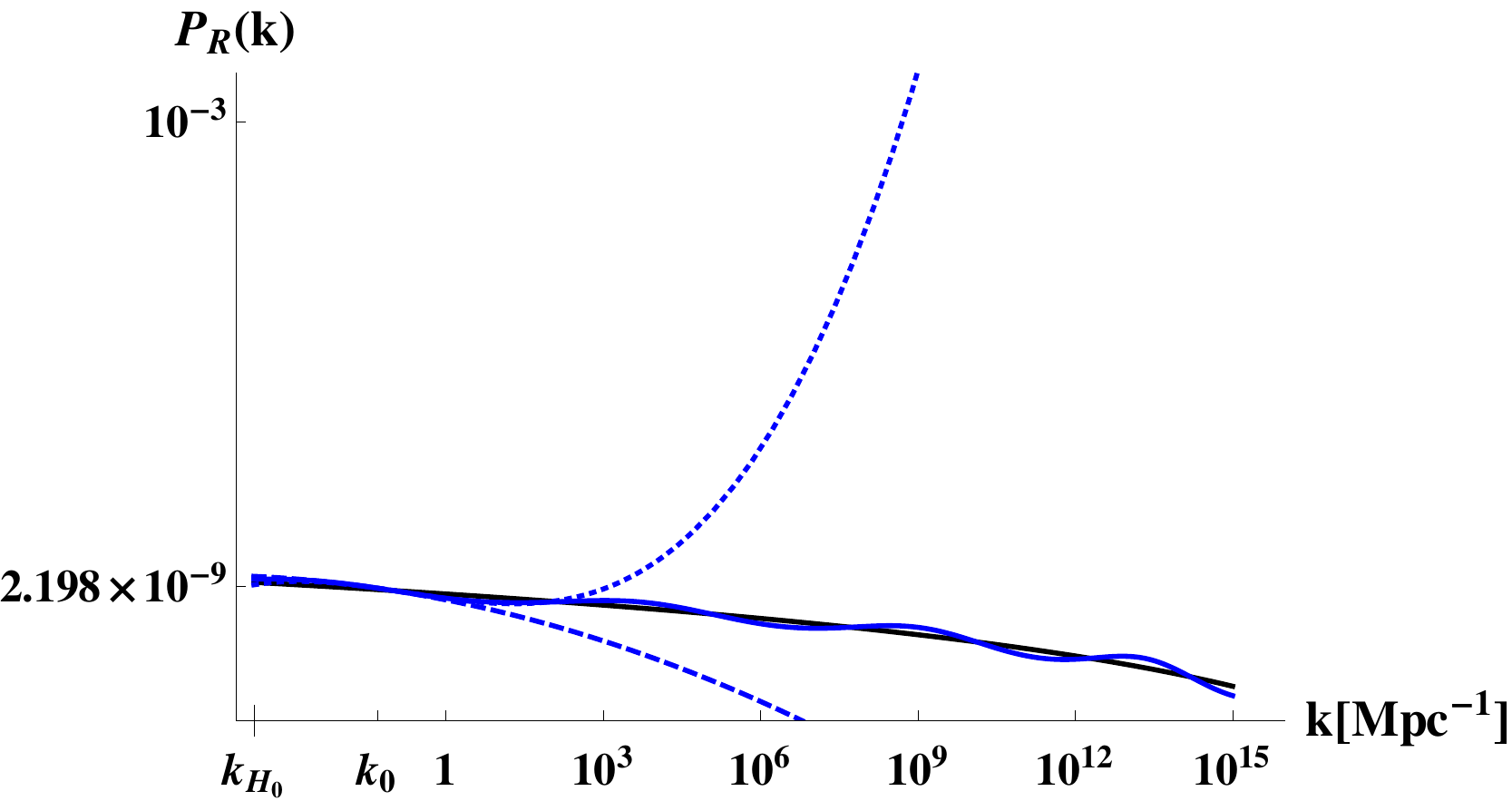}}
\caption{Power spectrum of curvature perturbations from the inflaton potential \eqref{model12} for $p=2$ (solid blue curve). Spectrum by including the running (of the running) is shown as the blue dashed (dotted) curve. Spectrum from a potential without any modulation is shown as the black line for comparison. Note that all the curves realize the observed value of the spectral index at the pivot scale $k_{0}=0.05$ Mpc$^{-1}$.}
\label{fig:chaotic2}
\end{figure}

\subsection{Hilltop/inflection point Inflation}

In another interesting class of potentials, the inflaton rolls away from an unstable equilibrium as in the first new inflationary models \cite{review, Albrecht, Linde1981}. The potential is given by
\be\label{model21}
V(\phi) \approx \lambda^4\left[1-\lp\dfrac{\phi}{\mu}\rp^p\right]\,,
\ee
where $\lambda,\,\mu$, and $p$ are positive constants.\footnote{This potential is unbounded from below for $\phi\rightarrow\infty$. There must be additional terms that prevent this. Here we follow the usual assumption that these terms do not affect the dynamics of inflation.} This potential is equivalent to the potential $V(\phi) = \lambda^4-\rho\,M_{\rm P}^{4-p}\dfrac{\phi^p}{p}$ in Ref.~\cite{Adshead}, which can be specialized to several distinct models, \eg, hilltop ($p=2$ or $p=4$) and inflection point ($p=3$). When $p$ is an integer greater than $2$, such a potential may be generated by the self--coupling of the inflaton at tree level. This model has been studied in detail for DM PBHs in Ref.~\cite{Encieh2}. The potential with an exponent of $p=2$ is in agreement with \p+WP+BAO joint $95\%$ C.L. contours for super-Planckian values of $\mu$, \ie, $\mu \gsim 9 \mpl$ \cite{Planck2}. The inflection point potential with $p=3$ lies outside the joint $95\%$ C.L. region for \p+WP+BAO data; the case with $p=4$ is also in tension with \p+WP+BAO but allowed within the joint $95\%$ C.L. region for $N \gsim 50$ \cite{Planck2}.

Adding a modulation term we have the following potential:
\be\label{model22}
V(\phi)=\lambda^4\left[1-\lp\dfrac{\phi}{\mu}\rp^p\right]+\Lambda^4\cos\lp\dfrac{\phi}{f}+\theta\rp\,,
\ee
where $\Lambda,\,f,$ and $\theta$ are constants. We consider the case that the dominant term is the leading one, $\lambda^4$. For $p>0$, the hierarchies hold among slow--roll parameters if $\phi \ll \mu$, so the spectral parameters are given by
\bea\label{model23}
n_s      &\simeq& 1-2\mpl^2\left[p(p-1)\dfrac{1}{\mu^2}\lp\dfrac{\phi}{\mu}\rp^{p-2}+\dfrac{1}{f^2}\lp\dfrac{\Lambda}{\lambda}\rp^{4}\cos\lp\dfrac{\phi}{f}+\theta\rp\right]\,,\\
\alpha_s &\simeq& -2p\mpl^4\left[p(p-1)(p-2)\dfrac{1}{\mu^4}\lp\dfrac{\phi}{\mu}\rp^{2(p-2)}-\dfrac{1}{\mu f^3}\lp\dfrac{\Lambda}{\lambda}\rp^{4}\lp\dfrac{\phi}{\mu}\rp^{p-1}\sin\lp\dfrac{\phi}{f}+\theta\rp\right]\,,\nn\\
\beta_s  &\simeq& -2p^2\mpl^6\left[2p(p-1)(p-2)^2\dfrac{1}{\mu^6}\lp\dfrac{\phi}{\mu}\rp^{3(p-2)}-\dfrac{1}{\mu^2 f^4}\lp\dfrac{\Lambda}{\lambda}\rp^{4}\lp\dfrac{\phi}{\mu}\rp^{2(p-1)}\cos\lp\dfrac{\phi}{f}+\theta\rp\right]\,.\nn
\eea
In the case at hand, the power spectrum is as follows:
\be\label{model24}
\mathcal{P}_{\mathcal{R}_c} \simeq \dfrac{\lambda}{12\pi^2 p^2 \mpl^6}\dfrac{\lambda^4\mu^{2P}}{\phi^{2(p-1)}}\left[1+\lp\dfrac{\Lambda}{\lambda}\rp^4\dfrac{\mu^p}{p f\phi^{p-1}}\sin\lp\dfrac{\phi}{f}+\theta\rp\right]^{-2}\,.
\ee
Here we study the inflection point $p=3$ and the hilltop $p=4$ potentials and we set the parameters of the models in Table~\ref{tab:2} such that the observable parameters satisfy the observational data from \p+WP+highL [eq.~\eqref{o6} data].
\begin{table}[h!]
\centering
\begin{tabular}{ | c | c | c | }
  \cline{2-3}
  \multicolumn{1}{c|}{}    & $p=3$                   & $p=4$                 \\\cline{1-3}
  $\lambda$                & $7.6\times10^{-3}$      & $4.9\times10^{-3}$    \\\cline{1-3}
  $\Lambda$                & $9.0\times10^{-4}$      & $8.5\times10^{-4}$    \\\cline{1-3}
	$\phi_{_0}$              & $6.9$                   & $5.5$                 \\\cline{1-3}
	$\mu$                    & $10.8$                  & $10.8$                \\\cline{1-3}
	$f$                      & $0.23$                  & $0.12$                \\\cline{1-3}
	$\theta$                 & $2.3$                   & $-2.3$                \\\cline{1-3}
\end{tabular}
\caption{Parameter values of the inflection point $p=3$ and the hilltop $p=4$ potentials where $\phi_{_0}$ is the value of the inflaton field when it leaves the {\it Planck} pivot scale. }
\label{tab:2}
\end{table}

\begin{figure}[h!]
\centering{\includegraphics[width=0.45\textwidth]{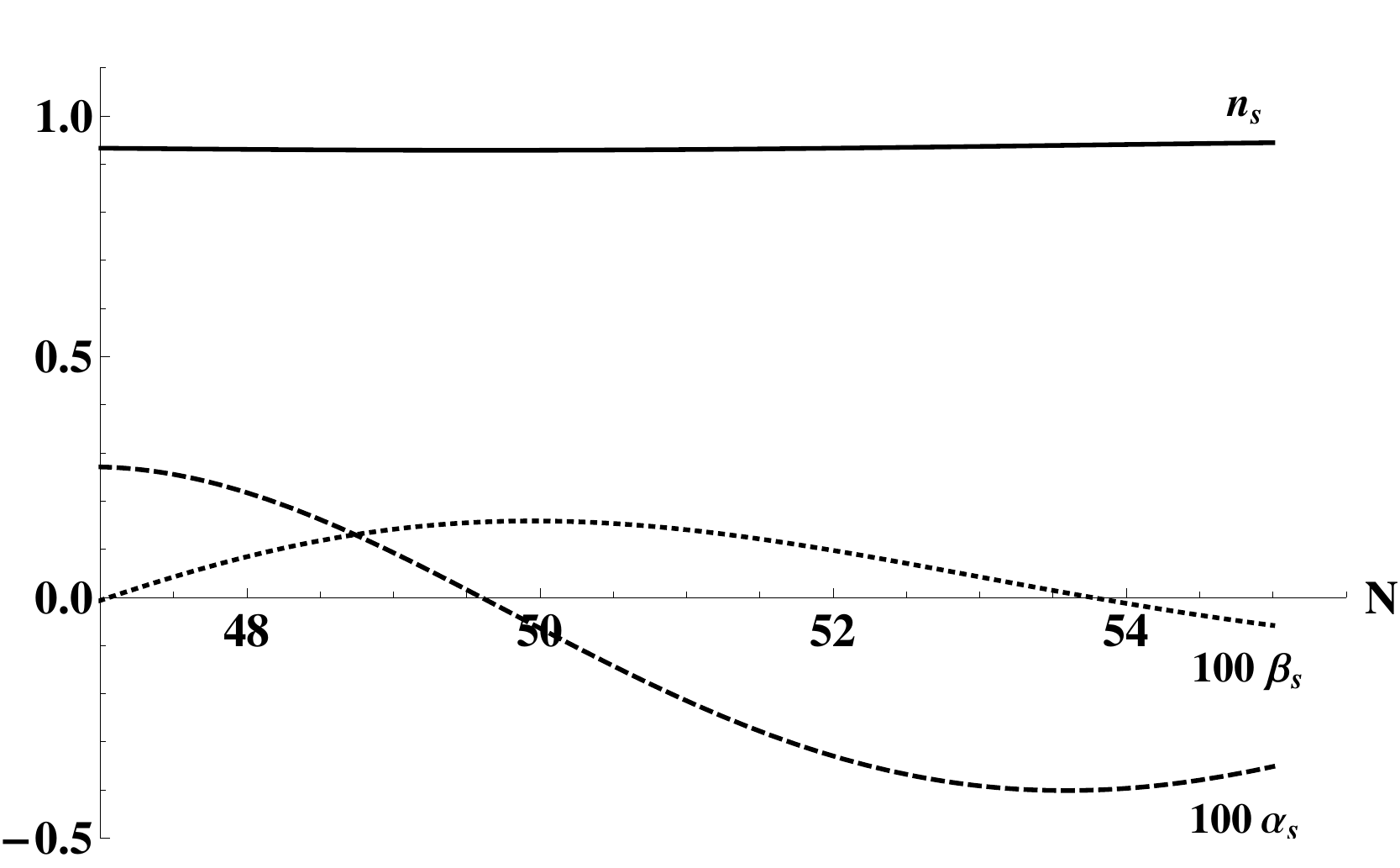}
\hspace{0.3cm}
\includegraphics[width=0.45\textwidth]{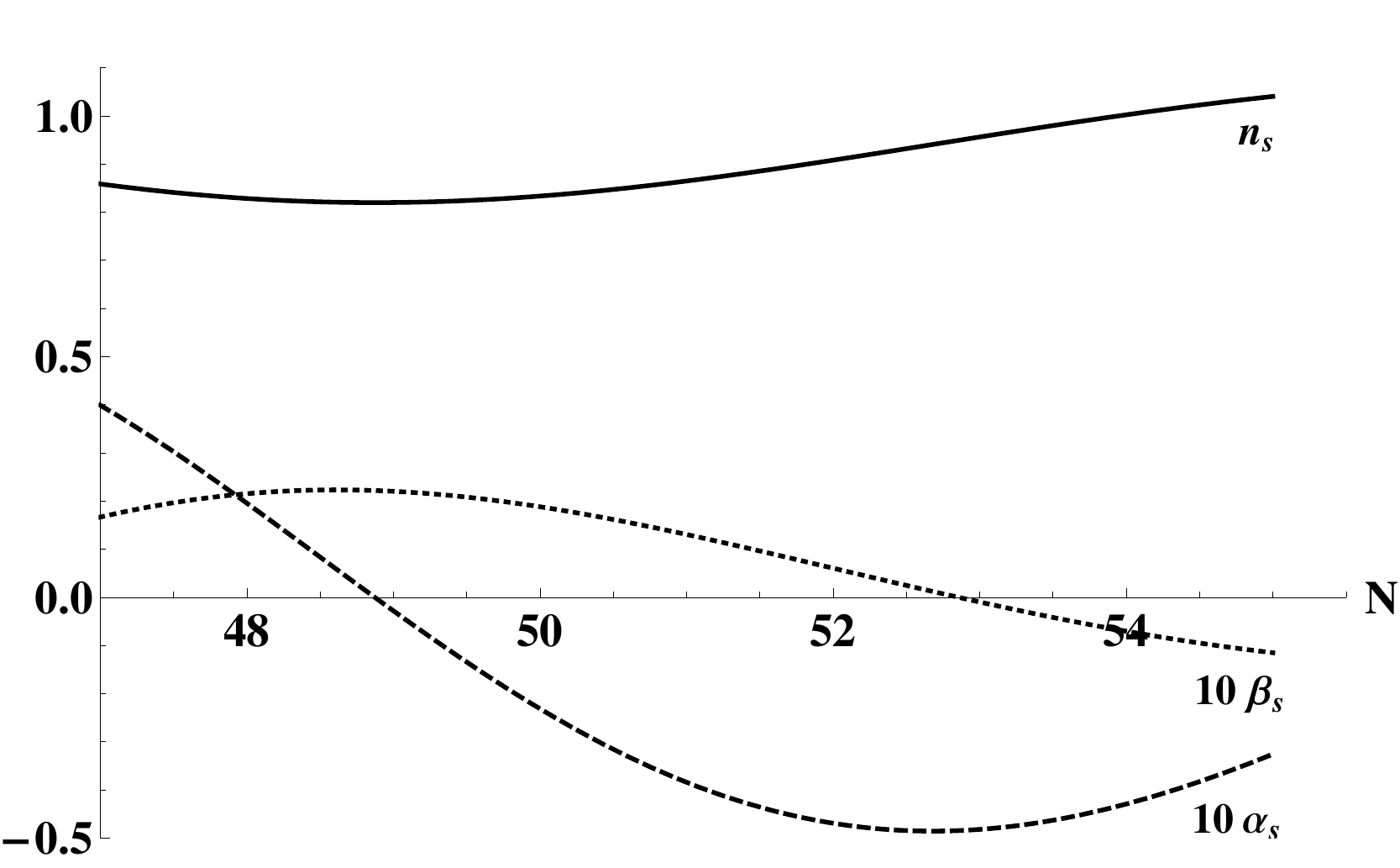}}
\caption{Oscillatory behavior of inflation parameters, $n_s$ (solid line), $\alpha_s$ (dashed line), and $\beta_s$ (dotted line) with respect to the number of $e$--folds for $p=3$ inflection point (left plot) and $p=4$ hilltop (right plot) potentials.}
\label{fig:par21}
\end{figure}

The oscillatory behavior of observable parameters (\ie, $n_s,\,\alpha_s$, and $\beta_s$) with respect to the number of $e$--folds is shown in Figure~\ref{fig:par21}.

\begin{figure}[h!]
\hspace*{-2.cm}
\vspace*{-1.5cm}
\includegraphics[width=1.\textwidth]{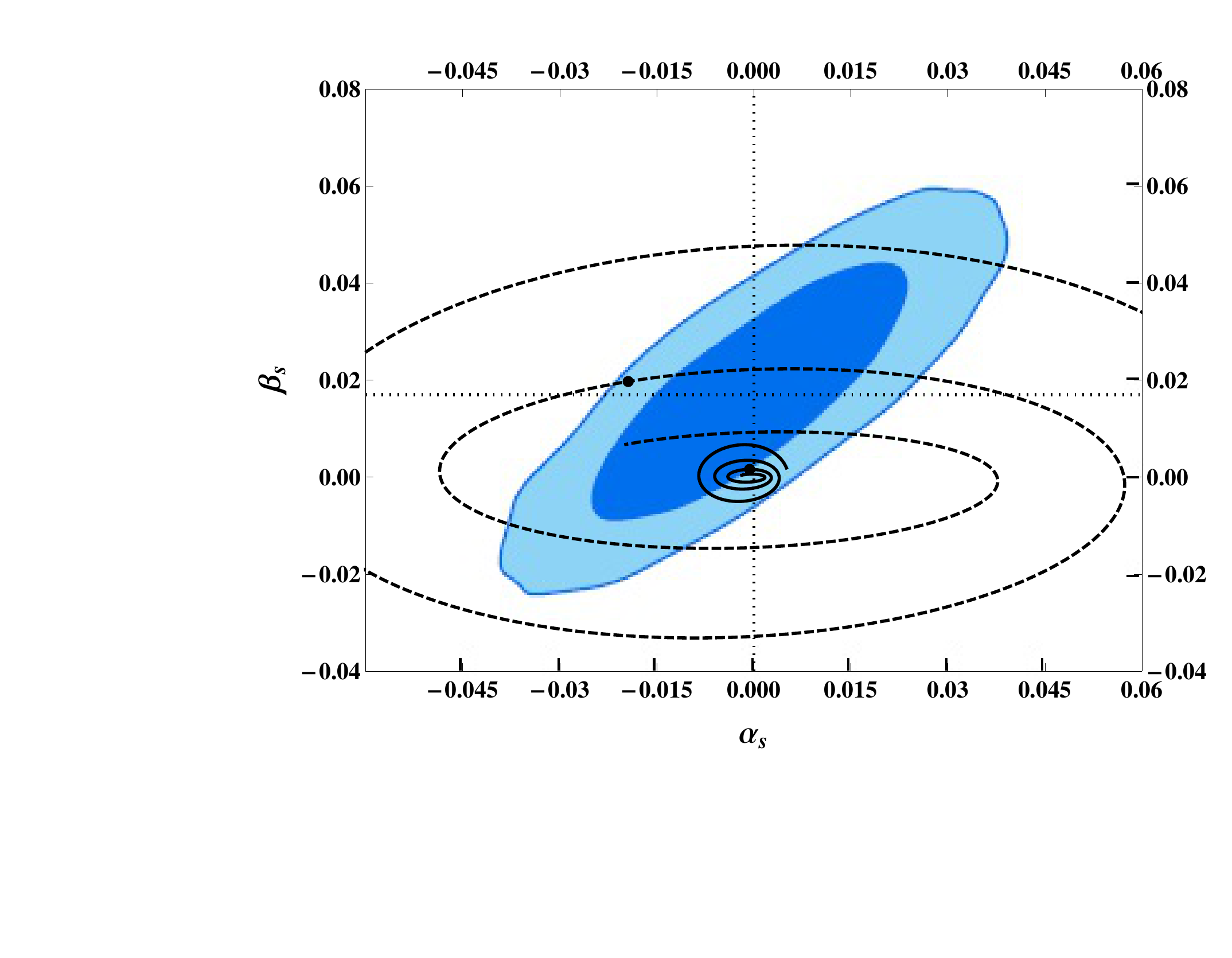}
\vspace*{-1.cm}
\caption{Variation of the running of the spectral index $\alpha_s$ and its derivative $\beta_s$ as inflaton proceeds along the $p=3$ inflection point (solid curve) and the $p=4$ hilltop (dashed curve) potentials with modulation. The dotted lines are the central values of the parameters as reported in \p+WP+BAO data and the curves start from 70 $e$--folds before the end of inflation (end point of innermost orbits) up to 30 $e$--folds before the end of inflation. Black points correspond to the pivot scale (about 50 $e$--folds before the end of inflation). The superimposed contour is marginalized joint $68\%$ and $95\%$ C.L. regions using \p+WP+BAO data \cite{Planck2}.}
\label{fig:par22}
\end{figure}

Since we are interested in the running of the running of the spectral index, we plot oscillating trajectories in the $\alpha_s-\beta_s$ plane for $p=3$ and $p=4$ potentials with modulation in Figure~\ref{fig:par22}. It is clear that in these inflationary models, the value of the $\alpha_s$ and its running are in the $2\,\sigma$ range of the \p+WP+BAO data, and the value of $\beta_s$ is positive, which is necessary for DM PBHs formation. Normalized at the pivot scale, one sees that although the $p=3$ inflection point potential was in tension with \p+WP+BAO data, by adding the modulated term, this model is consistent with the data. Although the $\beta_s$ is positive in these models, as mentioned before, this condition is not sufficient for DM PBHs formation. To clarify this point further, by the parameter sets in Table~\ref{tab:2}, we calculate the value of the spectral index at the scale relevant for (long--lived) PBHs formation. The results are as follows:
\bea
p=3 \quad &\Rightarrow& n_s(k_{\rm PBH})\simeq 1.23 < 1.37\,,\nn\\
p=4 \quad &\Rightarrow& n_s(k_{\rm PBH})\simeq 4.91 > 1.37\,.\nn
\eea

It is clear that in the $p=4$ hilltop model, the value of the spectral index at the DM PBHs scale is larger than $1.37$; therefore, PBHs can form in this model. However, although in the $p=3$ inflection point model, the running of the running of the spectral index $\beta_s$ is positive, the DM PBHs cannot form in this case. Therefore, the positive value of $\beta_s$ is only the necessary condition for DM PBHs formation, and for their formation, the power spectrum should reach $\sim 10^{-3}$ at the relevant scales. These results can also be seen by computing the power spectrum in the CMB and the scale of DM PBHs formation range.

\begin{figure}[h!]
\centering{\includegraphics[width=0.48\textwidth]{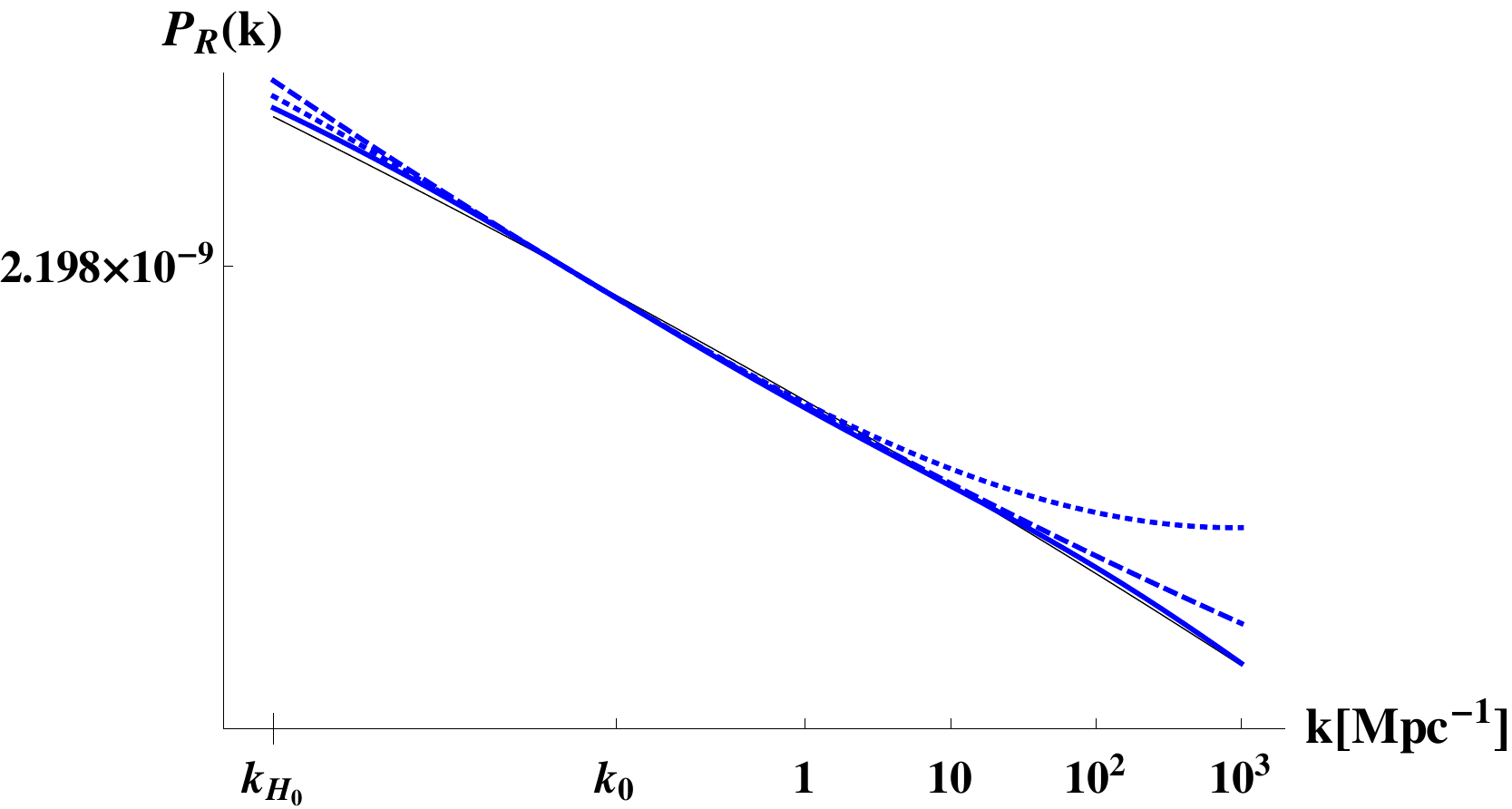}
\hspace{0.1cm}
\includegraphics[width=0.48\textwidth]{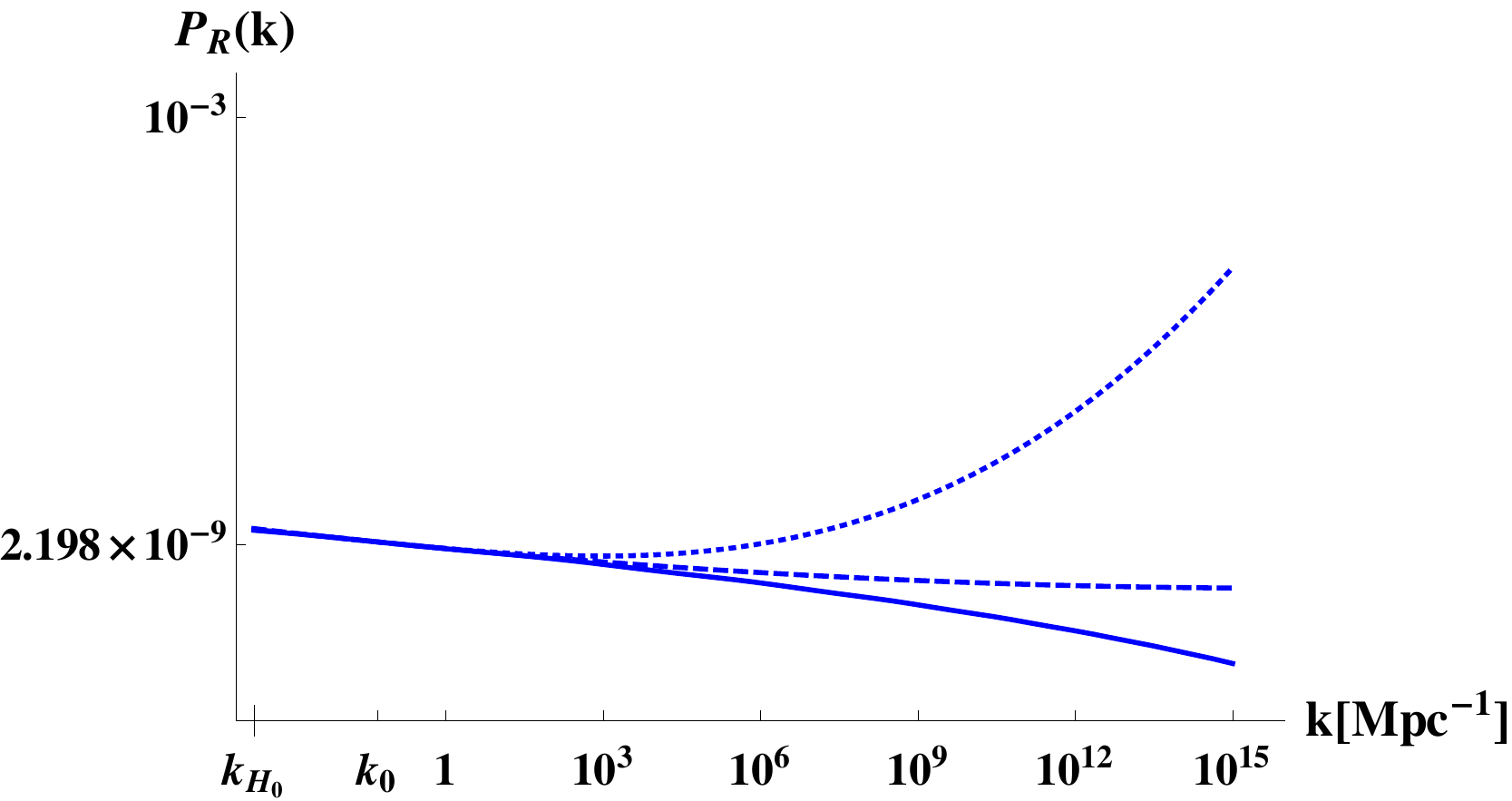}}
\caption{Curvature perturbation spectrum from the $p=3$ modulated inflection point inflaton potential with spectral index (solid blue curve). Spectrum by including the running (of the running) is shown as the blue dashed (dotted) curve. Spectrum from a potential without any modulation is shown as the black line, but because of small modulation it is not distinguishable from the power spectrum with modulation (solid blue curve). Note that all the curves realize the observed value of the spectral index at the pivot scale $k_{0}=0.05$ Mpc$^{-1}$.}
\label{fig:hilltop1}
\end{figure}

\begin{figure}[h!]
\centering{\includegraphics[width=0.48\textwidth]{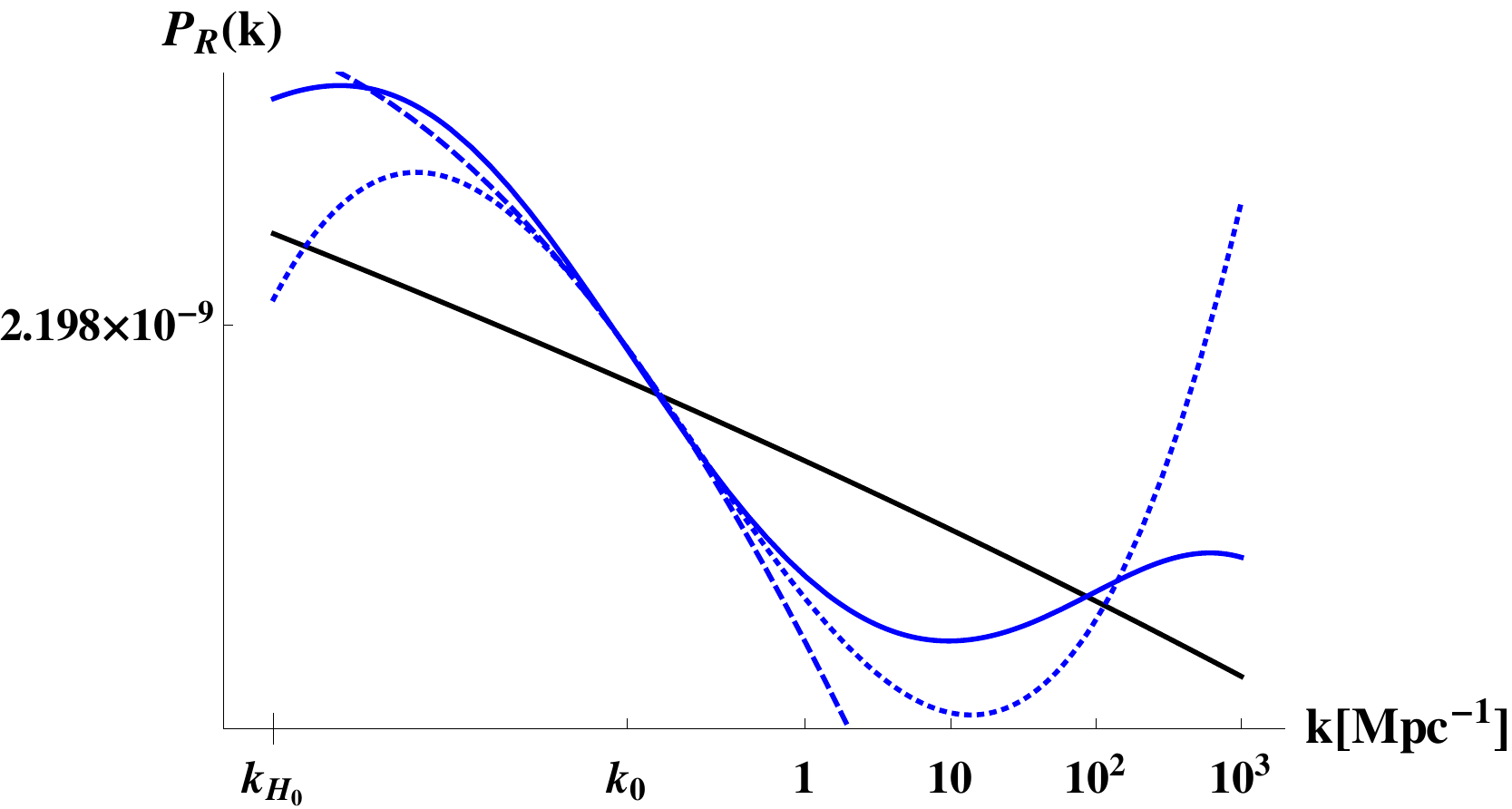}
\hspace{0.1cm}
\includegraphics[width=0.48\textwidth]{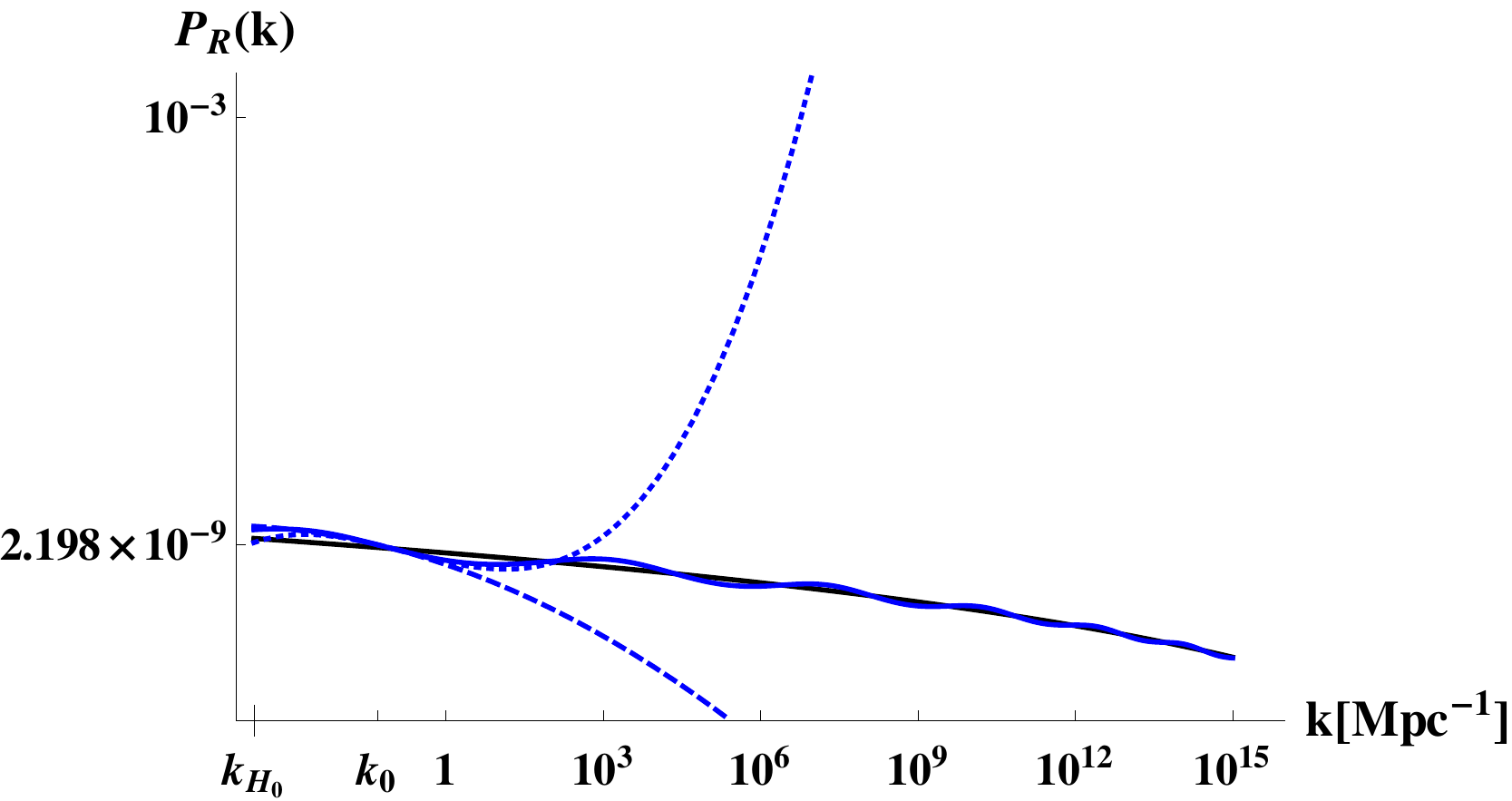}}
\caption{Curvature perturbation spectrum from the $p=4$ modulated hilltop inflaton potential with spectral index (solid blue curve). Spectrum by including the running (of the running) is shown as the blue dashed (dotted) curve. Spectrum from a potential without any modulation is shown as the black line for comparison. Note that all the curves realize the observed value of the spectral index at the pivot scale $k_{0}=0.05$ Mpc$^{-1}$.}
\label{fig:hilltop2}
\end{figure}

As the previous model, we compute the power spectrum by solving numerically the inflaton equation of motion and the Friedmann equation. The results are shown in figures~\ref{fig:hilltop1} and \ref{fig:hilltop2} for the inflection point and hilltop models, respectively.

Note that in the $p=3$ modulated inflection point inflation model, since the modulations are too small the modulated (blue solid curve) and nonmodulated (black solid line) power spectra are not distinguishable in Figure~\ref{fig:hilltop1}. From the right plot of Figure~\ref{fig:hilltop1}, it is clear that in this model, even by including the running of the running of the spectral index (dotted blue curve), the power spectrum cannot reach the required value of $\mathcal{O}(10^{-3})$ at the scale of $10^{15}$ g PBHs formation, \ie, $k_{\rm PBH}\simeq10^{15}$ Mpc$^{-1}$. Hence, the long--lived PBHs cannot form in the modulated inflection point inflation model.

In Ref.~\cite{Encieh2}, the PBHs formation in the $p=4$ hilltop inflation model without a modulated term has been studied. In this model without modulation, it was not possible to get the positive value of order $10^{-3}$ for $\beta_s$ to form PBHs. However, as shown in Figure~\ref{fig:par22}, the modulated hilltop inflation model is still consistent with the recent observational data and even PBHs can form in this model (right plot of Figure~\ref{fig:hilltop2}).

\section{Summary and Conclusions}

In this paper, we studied the possibility of primordial black holes formation in the chaotic, inflection point and hilltop inflationary models with superimposed periodic modulations after {\it Planck} data by assuming the slow--roll approximations. Although quadratic (chaotic) inflation and inflection point inflationary models are in tension with \p+WP+highL data (\ie, the spectral index lies outside the $95\%$ C.L. region in the $n_s-r$ plane), when a modulated term is added, we have seen that these models are consistent with the recent data. We showed that the tension can be relaxed in these modified models, and the $2\sigma$ range of values of the inflation parameters, \ie, the spectral index $n_s$, its first and second derivatives, $\alpha_s$ and $\beta_s$, respectively, can be realized. What differs from the previous studies using potentials without the modulated term is that the spectral index and its running are not localized, and change sign during the inflation. We focused on a case in which a sizable (positive) running of the running of the spectral index - which is crucial for dark matter PBHs formation - is generated due to the additional modulated term. In our analysis, we also assumed that the period of oscillations is large enough to encompass the observed scales by \p+WP+highL data set.

It was shown \cite{Encieh1} that for the formation of PBHs with mass larger than $10^{15}$ g that survive the Hawking radiation \cite{Hawking} and could be candidate for dark matter, the spectral index at scale of PBHs formation $k_{\rm PBH}$ is about $1.37$, which is completely independent of any inflationary model. This spectral index is much above the value measured at much larger length scales in the CMB. Therefore, PBH formation requires significant positive running of the spectral index when $k$ is increased. We compared this with the values of the spectral index and its running derived from current data on large scale structure. These include analysis of CMB anisotropies from the {\it Planck}, WMAP nine-years polarization, and as well as data from ACT and SPT collaborations, \ie, \p+WP+highL data set. At the pivot scale of this data set, one finds $n_s(k_0)=0.959$ as the central value. The first derivative $\alpha_s(k_0)$ would then need to exceed $0.024$ if it alone were responsible for the required increase of the spectral index; this is more than $4.5\sigma$ above the current central value of this quantity. Therefore, the positive value of the second derivative of the spectral index $\beta_s$ at the pivot scale of the observational data is required. However, this condition is a necessary but not a sufficient condition, since for PBHs formation, the power spectrum should reach $\sim10^{-3}$ at the relevant scales. Currently, the running of the running of the spectral index constrained by {\it Planck}. Our study showed that the nonproduction of DM PBHs put a stronger upper bound on the value of $\beta_s$ [see eqs.~\eqref{bound1} and \eqref{bound2}].

In this paper, we have studied the possibility of periodic superimposed modulation to the (linear and quadratic) chaotic, inflection point ($p=3$), and hilltop ($p=4$) inflaton potentials. We showed that although in the inflection point model, the running of the running of the spectral index is positive, but since the power spectrum of the density perturbations cannot reach the required value in the relevant scale of PBHs formation, long--lived PBHs cannot form in this model. However, in the $p=1$ and $p=2$ chaotic and hilltop inflationary models, it is possible to form PBHs.
 
It is worth mentioning that due to the oscillation term in superimposed inflationary model, the spectral index $n_s$, its running $\alpha_s$, and its running of running $\beta_s$ change sign. Such behavior results in distinct feature in the power spectrum of the curvature perturbations at small scales. Such characteristic behavior of the density perturbation may be seen by the future 21 cm observations \cite{21cm}. The potential detection of such oscillatory feature in the power spectrum by future observations was studied in Ref.~\cite{Hamann}, with the main focus on oscillations with (much) shorter period. In this work, we took the phenomenological approach where we explored the implications of a modulating term in inflationary models. It would be interesting to derive such a term from a fundamental theory. We leave this question for a future work.

\section*{Acknowledgments}
E.E. would like to thank I.~Zavala for useful comments on the draft. E.E. acknowledges partial support from the European Union FP7 ITN INVISIBLES (Marie Curie Actions, PITN-GA-2011-289442). E.E. is also grateful to the hospitality of the Abdus Salam ICTP where a part of this work was completed.

\end{document}